\documentclass[twocolumn,reprint,aps,prd,nofootinbib,floatfix,superscriptaddress]{revtex4-1}
\usepackage[utf8]{inputenc}
\usepackage[utf8]{inputenc}
\usepackage{amsmath}
\usepackage{amsfonts}
\usepackage{amssymb}
\usepackage[left=2cm,right=2cm,top=2cm,bottom=2cm]{geometry}
\usepackage{braket}
\usepackage{dsfont}
\usepackage{hyperref}

\usepackage{graphicx}
\usepackage{tikz}
\usetikzlibrary{arrows,decorations.pathmorphing,backgrounds,positioning,fit,petri,shapes}
\usepackage{lipsum}

\newcommand\blfootnote[1]{%
  \begingroup
  \renewcommand\thefootnote{}\footnote{#1}%
  \addtocounter{footnote}{-1}%
  \endgroup
}

\usepackage{multirow}
\begin{document}

\title{$x$-cut Cosmic Shear: Optimally Removing Sensitivity to Baryonic and Nonlinear Physics with an Application to the Dark Energy Survey Year 1 Shear Data}

\begin{abstract}

We present a new method, called $x$-cut cosmic shear, which optimally removes sensitivity to poorly modeled scales from the two-point cosmic shear signal. We show that the $x$-cut cosmic shear covariance matrix can be computed from the correlation function covariance matrix in a few minutes, enabling a likelihood analysis at virtually no additional computational cost. Further we show how to generalize $x$-cut cosmic shear to galaxy-galaxy lensing. Performing an $x$-cut cosmic shear analysis of the Dark Energy Survey Year 1 (DESY1) shear data, we reduce the error on $S_8= \sigma_8 (\Omega_m / 0.3) ^ {0.5}$ by $32 \%$ relative to a correlation function analysis with the same priors and angular scale cut criterion, while showing our constraints are robust to different baryonic feedback models. Largely driven by information at small angular scales, our result, $S_8= 0.734 \pm 0.026$, yields a $2.6 \sigma$ tension with the Planck Legacy analysis of the cosmic microwave background. As well as alleviating baryonic modelling uncertainties, our method can be used to optimally constrain a large number of theories of modified gravity where computational limitations make it infeasible to model the power spectrum down to extremely small scales. The key parts of our code are made publicly available.\footnote{\hyperlink{https://github.com/pltaylor16/x-cut}{https://github.com/pltaylor16/x-cut}.}
\end{abstract}

\author{Peter L.~Taylor}
\email{peterllewelyntaylor@gmail.com}
\affiliation{Jet Propulsion Laboratory, California Institute of Technology, 4800 Oak Grove Drive, Pasadena, CA 91109, USA}
\blfootnote{© 2020. All rights reserved.} 
\author{Francis Bernardeau}
\affiliation{UPMC - CNRS, UMR7095, Institut d’Astrophysique de Paris, F-75014, Paris, France}
\affiliation{CEA - CNRS, URA 2306, Institut de Physique Th$\acute{e}$orique, F-91191 Gif-sur-Yvette, France}
\author{Eric Huff}
\affiliation{Jet Propulsion Laboratory, California Institute of Technology, 4800 Oak Grove Drive, Pasadena, CA 91109, USA}

\maketitle

\section{Introduction}
Over the coming decade weak lensing will provide some of the tightest constraints on the mass of the neutrino and the dark energy equation of state. Data from {\it Stage IV} experiments including Euclid\footnote{\url{http://euclid-ec.org}} \cite{laureijs2010euclid}, the Nancy Grace Roman Space Telescope\footnote{\url{https://www.nasa.gov/wfirst}} \cite{spergel2015wide} and the Rubin Observatory\footnote{\url{https://www.lsst.org}} will revolutionize the field, increasing the number of observed galaxies by more than an order of magnitude.  Extracting cosmological information from these next generation surveys in a precise and unbiased way presents a formidable challenge. To take advantage of this data, methodologies must first be honed on current state-of-the-art {\it Stage III} surveys which include the Dark Energy Survey (DES)~\cite{troxel2018dark}, the Kilo-Degree Survey (KiDS)~\cite{hildebrandt2016kids} and the Subaru Hyper Suprime-Cam Lensing Survey~\cite{hikage2018cosmology}.
\par One particular challenge is dealing with modelling uncertainties at small scales. Cosmic shear is sensitive to poorly modelled scales -- down to $7 \ h {\rm Mpc} ^ {-1}$~\cite{taylor2018preparing} in the matter power spectrum and sub-megaparsec scales in configuration space. While it is not yet possible to derive accurate analytic predictions for these scales~\cite{osato2019perturbation}, one can model nonlinear structure growth using an emulator~\cite{lawrence2010coyote, euclid2019euclid, winther2019emulators} (or halo model~\cite{mead2015accurate}) trained (calibrated) on a suite of $\mathcal{O} (100)$~\cite{euclid2019euclid} high resolution N-body simulations, run over cosmological parameter space. 
\par Simulations must be run with more than a trillion particles to meet the percent-level matter power spectrum accuracy requirements of upcoming surveys~\cite{schneider2016matter}. Emulators and halo models with varying degrees of accuracy have been trained for Lambda cold dark matter (LCDM)~\cite{lawrence2010coyote, mead2015accurate}, $w_0$ cosmologies~\cite{euclid2019euclid} and a small subset of theories of popular theories of modified gravity (see e.g~\cite{winther2019emulators, bose2020road}). However, to test all theories of modified gravity (see~\cite{amendola2018cosmology} for a review), without throwing away information, we would need a suite of high resolution simulations for each theory. This sets a formidable and potentially unachievable task.
\par Baryonic physics further complicates the problem of accurately modelling structure growth at small scales. If not accounted for correctly, baryonic feedback will significantly bias cosmological parameter constraints~\cite{semboloni2011quantifying, Huang:2018wpy}. To make matters worse, the impact of baryonic processes can not be extracted from high resolution N-body simulations. Discrepancies between different `subgrid prescriptions' would have a large impact on parameter constraints~\cite{Huang:2018wpy}. 
\par Several methods to mitigate nonlinear and baryonic modelling uncertainties at small scales have been proposed. Some approaches are:
\begin{itemize}
\item{Taking na\"ive angular scale cuts. This corresponds to cutting large angular wave modes (small angular scales), $\ell$, from the lensing power spectrum, $C_\ell$, in harmonic space or small angles, $\theta$, from the two-point correlation function, $\xi_\pm (\theta)$, in configuration space. This method is always employed, whether explicitly or not, since no analysis uses all angular scales between zero and infinity. Used appropriately this technique yields unbiased yet imprecise parameter constraints.} 
\item{Reweighting the data with a carefully constructed transform before taking scale cuts. Complete Orthogonal Sets of E/B-Integrals (COSEBIs)~\cite{asgari2020kids+, asgari2019dark, schneider2010cosebis} and $k$-cut cosmic shear~\cite{taylor2018k-cut} both fall into this category. For COSEBIs a new data vector is computed as an integral transform of correlation functions before taking a cut in discrete wavenumber. Meanwhile in $k$-cut cosmic shear, the lensing power spectrum is reweighted to make the relationship between angular and physical scales more precise, before taking an angular scale cut. This paper is concerned with the configuration space analog of $k$-cut cosmic shear. Reweighted statistics can be used in combination with other baryonic mitigation strategies listed below.}
\item{Performing a principal component analysis to selectively remove linear combinations of the data vector which are most severely impacted by baryonic physics~\cite{eifler2015accounting, huang2018modeling}.}
\item{Using a physically motivated halo model~\cite{mead2015accurate} or fitting formula~\cite{harnois2015baryons} to capture the baryonic feedback which can then be marginalized out during a likelihood analysis. This is the approach used in~\cite{hikage2018cosmology, hikage2018cosmology}.}
\item{Calibrating physically motivated halo models on external observations~\cite{debackere2020impact, van2020exploring}. This will require additional targeted observations with dedicated telescope time.}
\end{itemize}
Ultimately some combination of the approaches listed is likely optimal.
\par In this paper we present a new method, which we refer to as $x$-cut cosmic shear, to cut sensitivity to poorly modeled scales -- while preserving useful information. 
\par This work is very similar to the recently proposed $k$-cut cosmic shear method presented in~\cite{taylor2018k-cut, deshpande2020accessing} (a similar method to remove sensitivity to small scales was given in~\cite{huterer2005nulling}). $k$-cut cosmic shear technique works by taking the Bernardeau-Nishimichi-Taruya (BNT)~\cite{bernardeau2014cosmic} transform (see~\cite{barthelemy2020nulling} for a further application of the BNT transform) which reorganizes the information making the relationship between the angular scale, $\ell$, and the structure scale, $k$, much clearer compared to standard cosmic shear power spectra -- before cutting large angular wavenumbers that correspond to small poorly modelled scales. $k$-cut cosmic shear has also recently been shown to minimize the impact of the reduced shear approximation~\cite{deshpande2020accessing} and forecasts for the Euclid survey have been provided in~\cite{taylor2020euclid}.
\par The key difference is that the method presented in this paper works in configuration space. This has several advantages over harmonic space. There is no need to deconvolve the mask which could lead to a loss of information~\cite{leistedt2013estimating}. Furthermore, it is more natural to take a cut in configuration space as non-linear structure which is restricted to the center of halos, and compact in configuration space, is spread over a large range of $k$-modes in Fourier space. 
\par We will show how to construct the $x$-cut statistic as a simple transformation of correlation functions. Further we show how to compute the $x$-cut covariance from the correlation function covariance matrix on a single CPU in a few minutes.  Performing an $x$-cut likelihood analysis of the DESY1 shear data, we will demonstrate that $x$-cut cosmic shear yields more precise constraints than a correlation function analysis while remaining robust to baryonic feedback modelling uncertainty.
\par The structure of this paper is the following. In Section~\ref{sec:formalism}, we review the standard cosmic shear formalism before presenting the $x$-cut cosmic shear method in Section~\ref{sec:x-cut}. A review of the data and public covariance used in the DESY1 cosmic shear analysis~\cite{troxel2018dark} (hereafter D18) is given in Section~\ref{sec:DESY1}. The baryonic feedback models used in this work are presented in Section~\ref{sec:baryons}. Our results and the $x$-cut cosmic shear parameter constraints are given in Section~\ref{sec:results} before discussing the future prospects for the method and concluding in Section~\ref{sec:conclusions}. 
\par We assume a LCDM cosmology with a free neutrino mass throughout. We compute the BNT transform and produce all figures assuming the baseline cosmology where: $(\Omega_m, h_0, \Omega_b, \Omega_\nu h_0 ^ 2, n_s, S_8)$ are taken to be (0.275, 0.7, 0.046, 0.004, 0.993, 0.78).

\section{Cosmic Shear Formalism} \label{sec:formalism}

\subsection{The Lensing Spectrum} 
The lensing power spectrum contains the two-point information of the shear field.
Under the Limber~\cite{loverdelimber, kitchinglimits}, spatially-flat universe~\cite{PLTtesting}, flat-sky~\cite{kitchinglimits}, reduced shear~\cite{deshpande2020euclid, dodelson2006reduced} and Zeldovich~\cite{kitchingunequal} approximations it is given by:
\begin{equation}
C_{GG}^{ij}(\ell) = \int_0^{\chi_H}d\chi \frac{q^i(\chi)q^j(\chi)}{\chi^2} P\left(\frac{\ell+1/2}{\chi},\chi\right),
\label{eq:CGG}
\end{equation}
where $\chi$ is the radial comoving distance, $P$ is the matter power spectrum, $\chi_H$ is the distance to the horizon, and $q(\chi)$ is the lensing efficiency kernel:
\begin{equation}
q^i(\chi) = \frac{3}{2}\Omega_m \left(\frac{H_0}{c}\right)^2 \frac{\chi}{a(\chi)} \int_{\chi}^{\chi_H} d\chi' n^i(\chi') \frac{\chi'-\chi}{\chi'},
\end{equation}
Here $H_0$ is the Hubble parameter, $\Omega_m$ is the fractional matter density parameter, $c$ is the speed of light, $a$ is the scale factor and $n^i(\chi')$ is probability distribution of the effective number density of galaxies as a function of comoving distance. In what follows we use {\tt CosmoSIS}~\cite{cosmosis} to compute the lensing spectrum using {\tt Camb}~\cite{camb} to generate the linear power spectrum and {\tt Halofit}~\cite{halofit} to generate the nonlinear power spectrum.

\subsection{Correlation Functions} 
Is it often more convenient to work in configuration space to avoid the need to deconvolve the survey mask using the pseudo-$C_\ell$ method~\cite{wandelt2001cosmic, brownpseudocl, alonso2018unified}, for example. In this case the two-point information is contained in the correlation functions which are defined as:
\begin{equation} \label{eqn:xiGG}
\xi_{\pm, {\rm GG}}^{ij}(\theta) = \frac{1}{2\pi}\int d\ell \,\ell \,C_{GG}^{ij}(\ell) \, J_{\pm}(\ell \theta), 
\end{equation}
where $J_{+}(\ell \theta)$ is the zeroth order Bessel function of the first kind and $J_{-}(\ell \theta)$ is the fourth order Bessel function of the first kind. We use {\tt Nicea}~\cite{kilbinger2009dark} to compute the correlation functions.

\subsection{Intrinsic Alignments} 
In addition to cosmic shear, the  ellipticity of galaxies is influenced by intrinsic alignments (IA)~\cite{hirata2004intrinsic} as galaxies tidally align with nearby dark matter halos. This leads to two additional contributions to the correlation functions: 
\begin{equation} \label{eqn:xifull}
{\xi}_{\pm}^{ij}(\theta) = \xi_{\pm,{\rm II}}^{ij}(\theta) +\xi_{\pm,{\rm GI}}^{ij}(\theta) +\xi_{\pm,{\rm GG}}^{ij}(\theta).
\end{equation}
An `II term' accounts for the intrinsic tidal alignment of galaxies around massive dark matter halos. Meanwhile the `GI term' accounts for the anti-correlation between tidally aligned galaxies at low redshifts and weakly lensed galaxies at high redshift.
\par As in D18, we follow the linear intrinsic alignment model originally given in~\cite{bridle2007dark} and used in~\cite{heymans2013cfhtlens}, allowing the amplitude of the alignments to vary as a function of redshift, as in~\cite{maccrann2015cosmic}. In this model the theoretical expression for II and GI correlation functions are:
\begin{equation}\label{eqn:xitheory}
\xi_{\pm,{\rm II/GI}}^{ij}(\theta) = \frac{1}{2\pi}\int d\ell \,\ell \,C_{\rm II/GI}^{ij}(\ell) \, J_{\pm}(\ell \theta) , 
\end{equation}
where the II spectrum, $C_{\rm II}^{ij}(\ell)$, is given by:
\begin{equation}
C_{\rm II}^{ij}(\ell) = \int_0^{\chi_{\rm H}} d \chi \, 
\frac{n^i(\chi)n^j(\chi)}{\chi^2} \, P_{\rm II} \left( \frac{\ell + 1/2}{\chi}, \chi \right),
\label{eqn:CII} 
\end{equation}
where the II matter power spectrum is:
\begin{equation} \label{eqn:PII}
P_{\rm II}(k,z) =  F^2(z) P(k,z)
\end{equation}
and
\begin{equation} \label{eqn:Fz} 
F(z) = - A_I \left[ \frac{1 + z}{1 + z_0}  \right] ^ \eta C_1 \rho_{\rm crit} \frac{\Omega_{\rm m}}{D(z)}.
\end{equation}
Here $\rho_{\rm crit}$ is the critical density of the Universe, $z_0$ is the median redshift of the survey, which for DESY1 is $z_0 = 0.62$, $D(z)$ is the growth factor and $C_1 = 5 \times 10^{-14} h^{-2} M_\odot^{-1} {\rm Mpc}^3$ is chosen so that the fiducial value of the intrinsic alignment amplitude, $A_I$, is unity~\cite{brown2002measurement}. 
Meanwhile the GI matter power spectrum is:
\begin{equation}
\begin{aligned}
C_{\rm GI}^{ij}(\ell) = \int_0^{\chi_{\rm H}} d \chi \, 
\frac{n^i(\chi)q^j(\chi) }{\chi^2}   P_{\rm GI} \left( \frac{\ell + 1/2}{\chi},\chi \right),
\label{eqn:CGI} 
\end{aligned}
\end{equation}
and the GI spectrum is:
\begin{equation} \label{eqn:PGI}
 P_{\rm GI}(k,z) =  F(z) P (k,z).
\end{equation}

\subsection{Systematics}
As in D18, we include two systematics which change the theoretical expectation of $\xi_\pm (\theta)$. For multiplicative biases $\{ m^i, m^j \}$ in bins $\{i,j \}$, the correlation functions become:
\begin{equation}
\xi_\pm^{ij} (\theta) \rightarrow (1 + m^i)(1 + m^j) \xi_\pm^{ij} (\theta).
\end{equation}
We also assume a linear photometric redshift bias, $\Delta z ^i$, so that the radial distribution function is shifted:
\begin{equation}
    n^i(z) \rightarrow n^i (z - \Delta z ^i).
\end{equation}
Multiplicative and redshift biases are treated as nuisance parameters and marginalized out at the end of the likelihood analysis. The choice of priors is informed by the measurement process. This is discussed further in Section~\ref{sec:DESY1}.

\section{$x$-cut Cosmic Shear} \label{sec:x-cut}

\subsection{Motivating the Bernardeau-Nishimichi-Taruya (BNT) Transformation}

 \begin{figure*}[!hbt]

\includegraphics[width = 5.5cm]{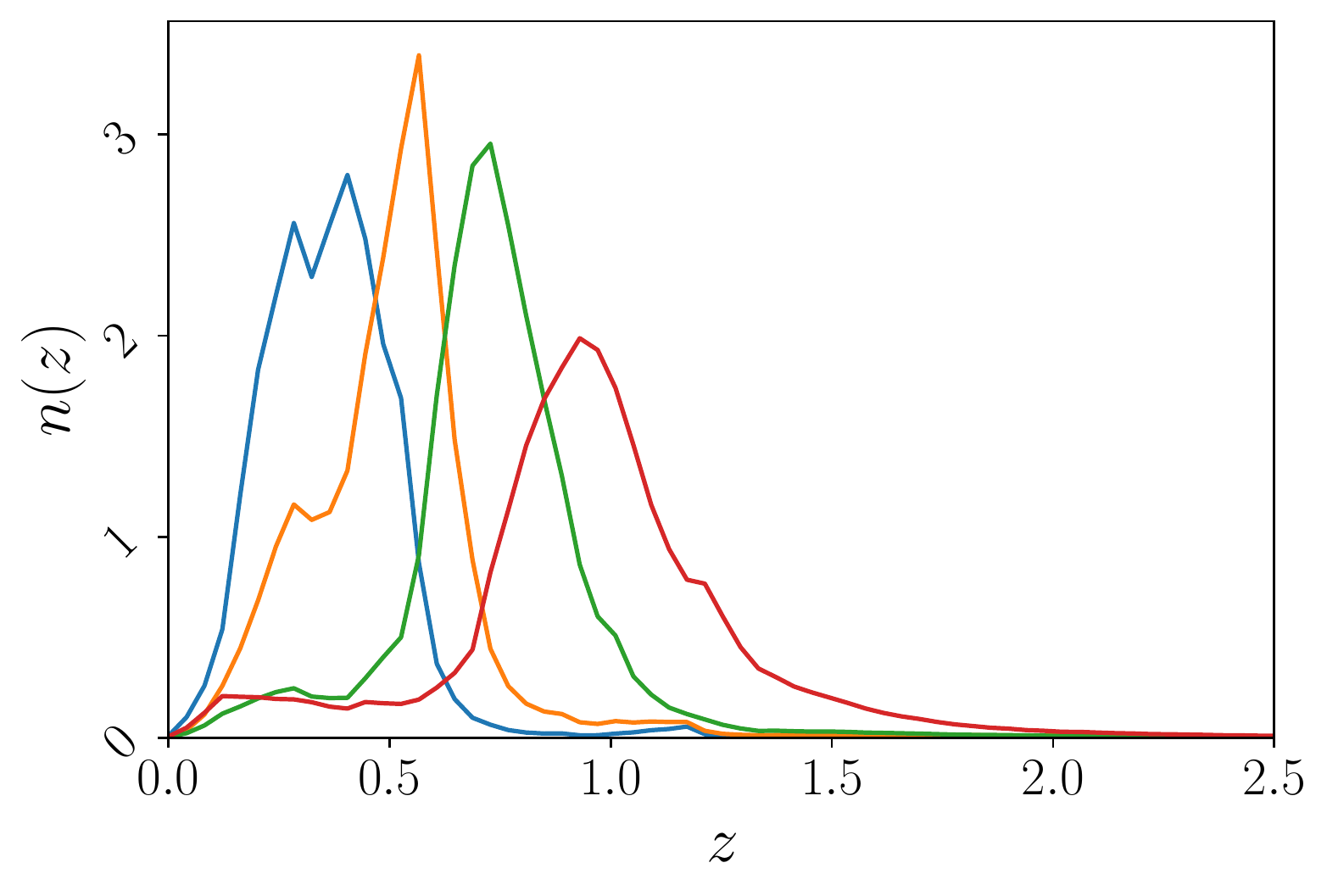}
\includegraphics[width = 5.5cm]{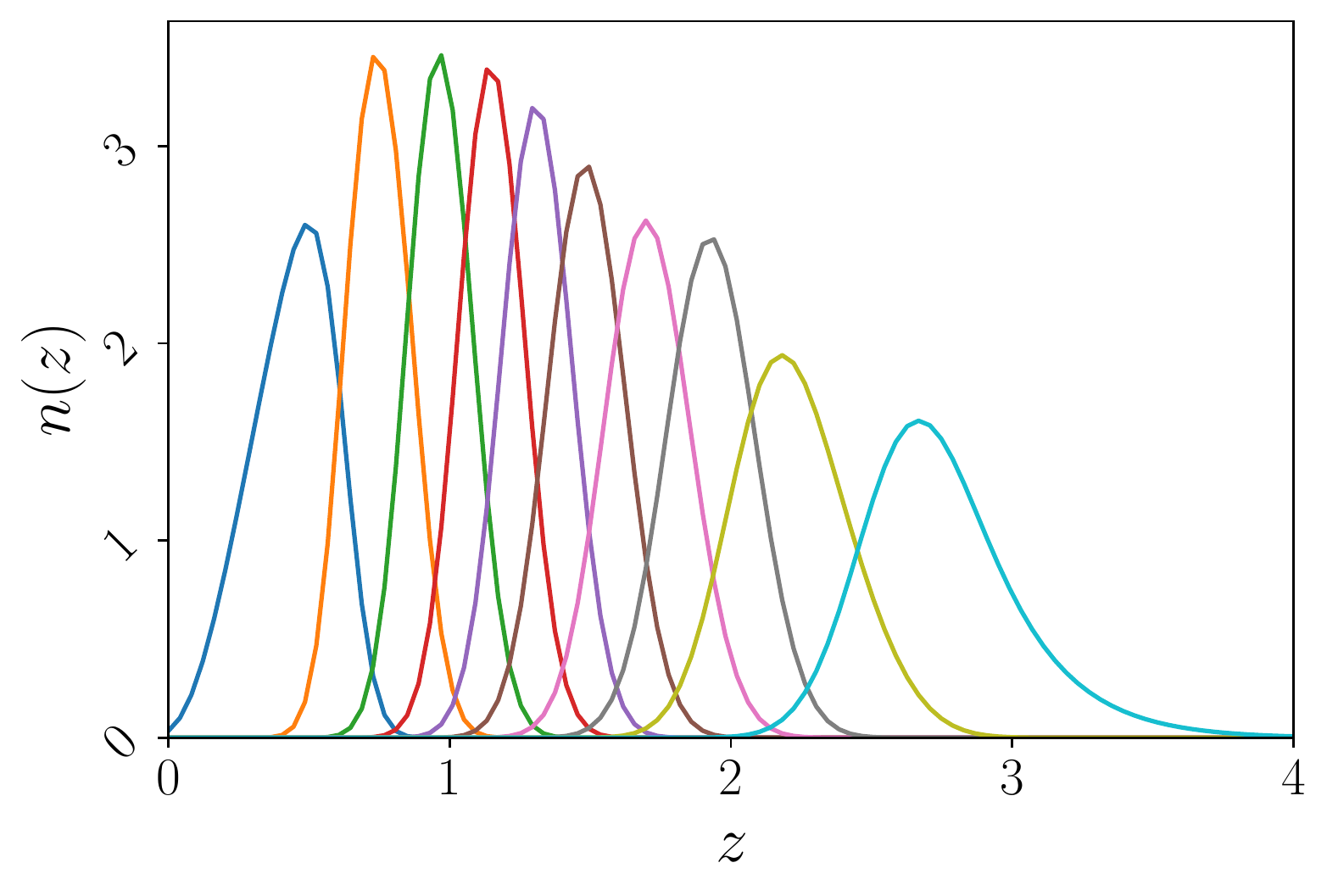}
\includegraphics[width = 5.5cm]{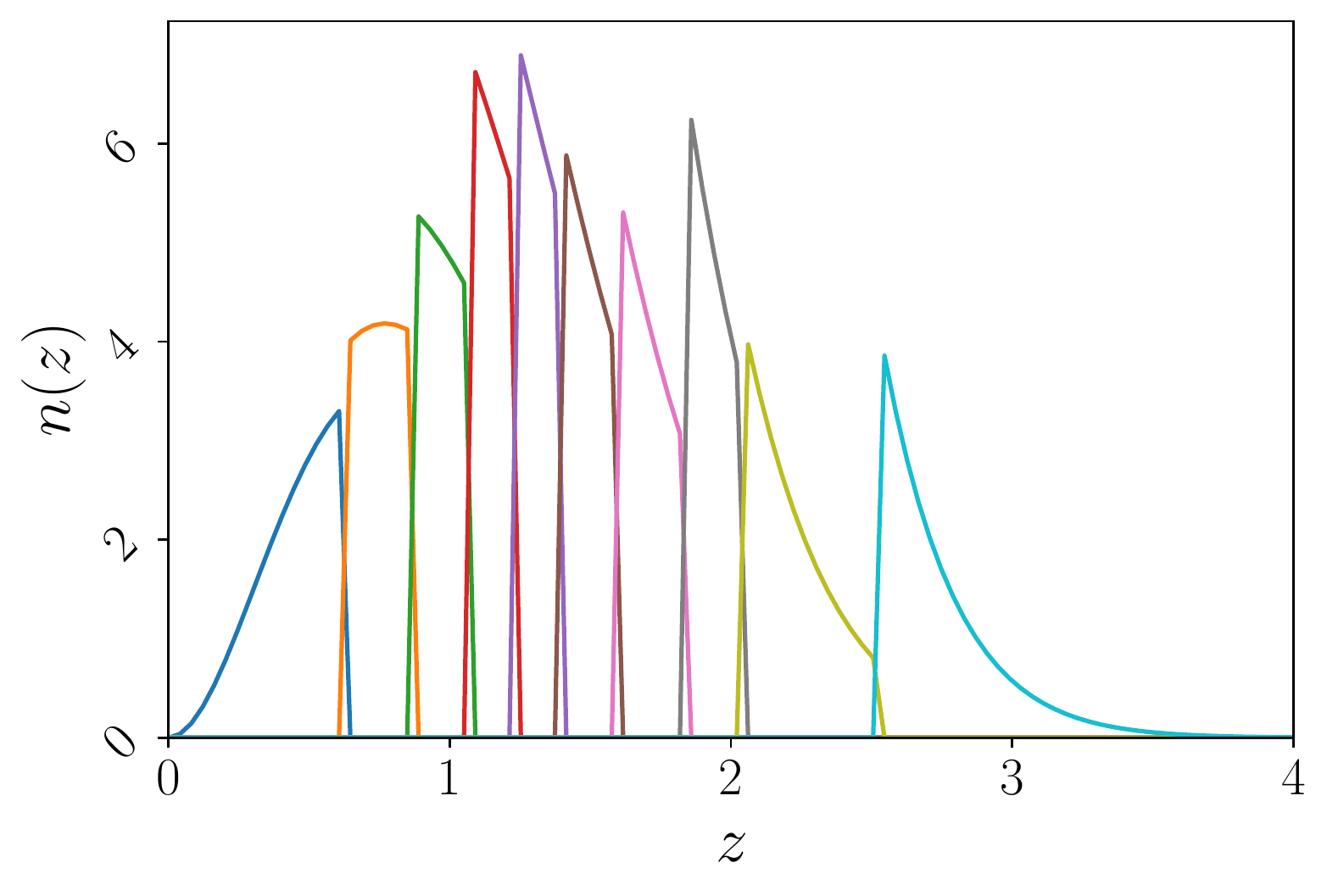}

\includegraphics[width = 5.5cm]{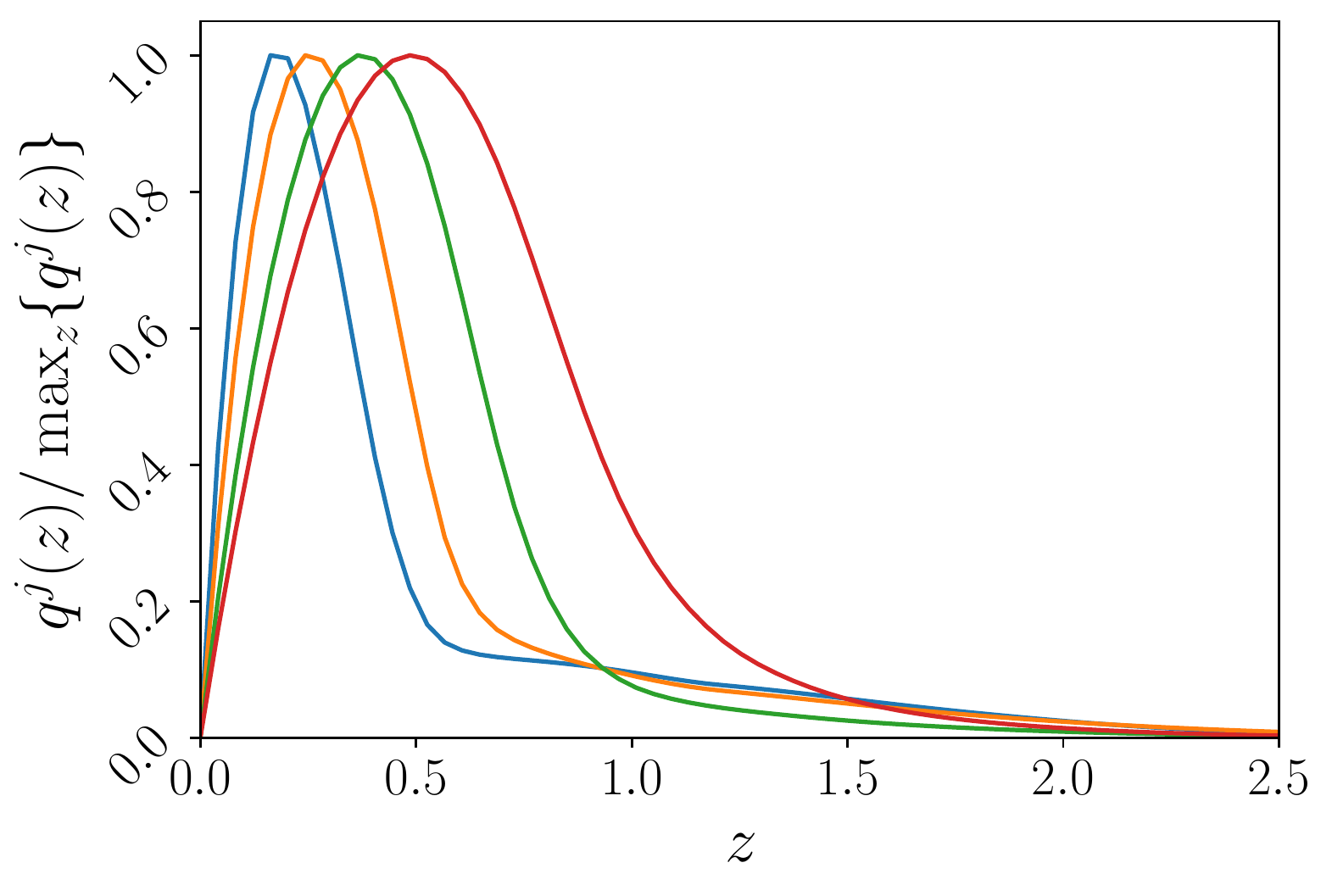}
\includegraphics[width = 5.5cm]{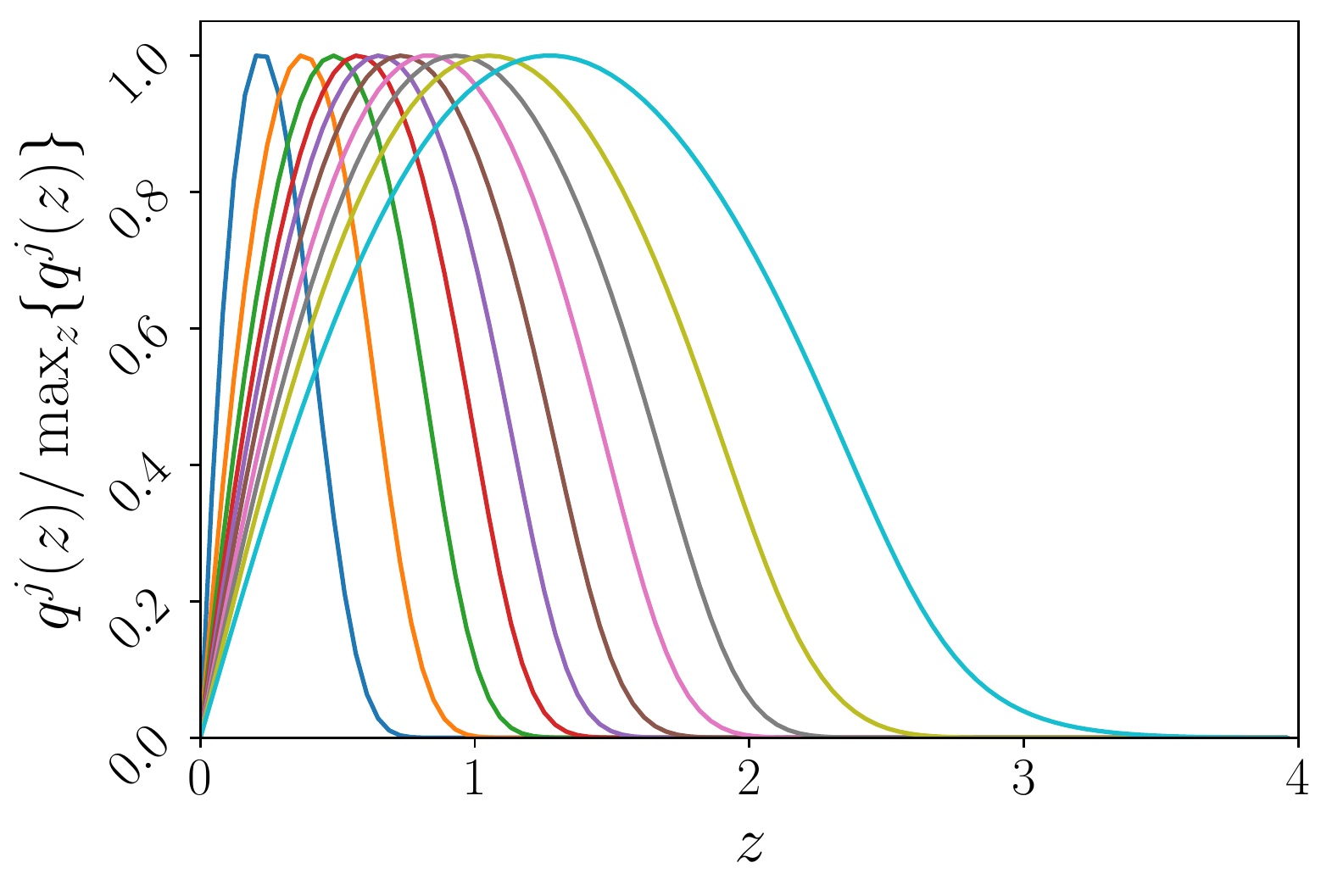}
\includegraphics[width = 5.5cm]{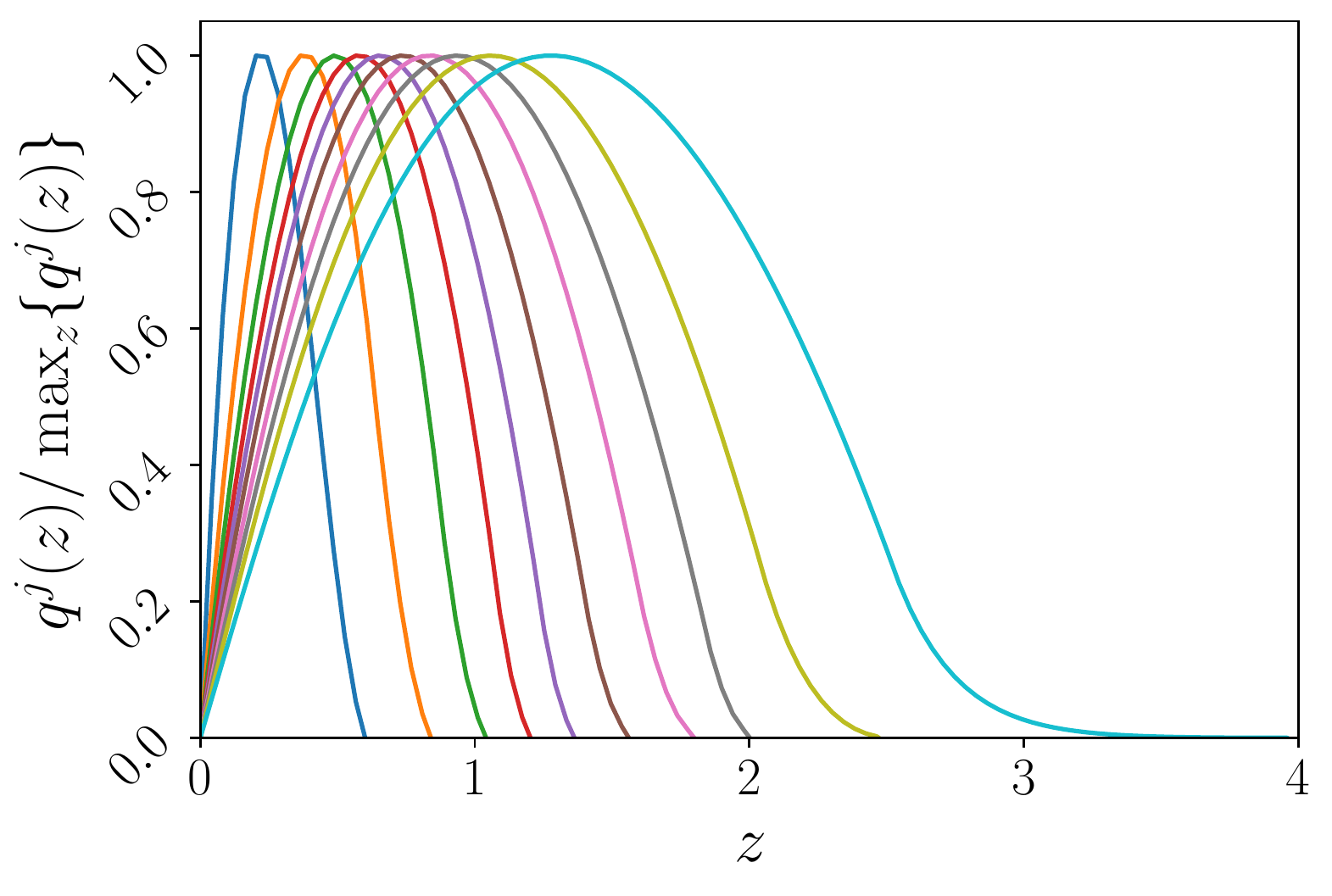}

\includegraphics[width = 5.5cm]{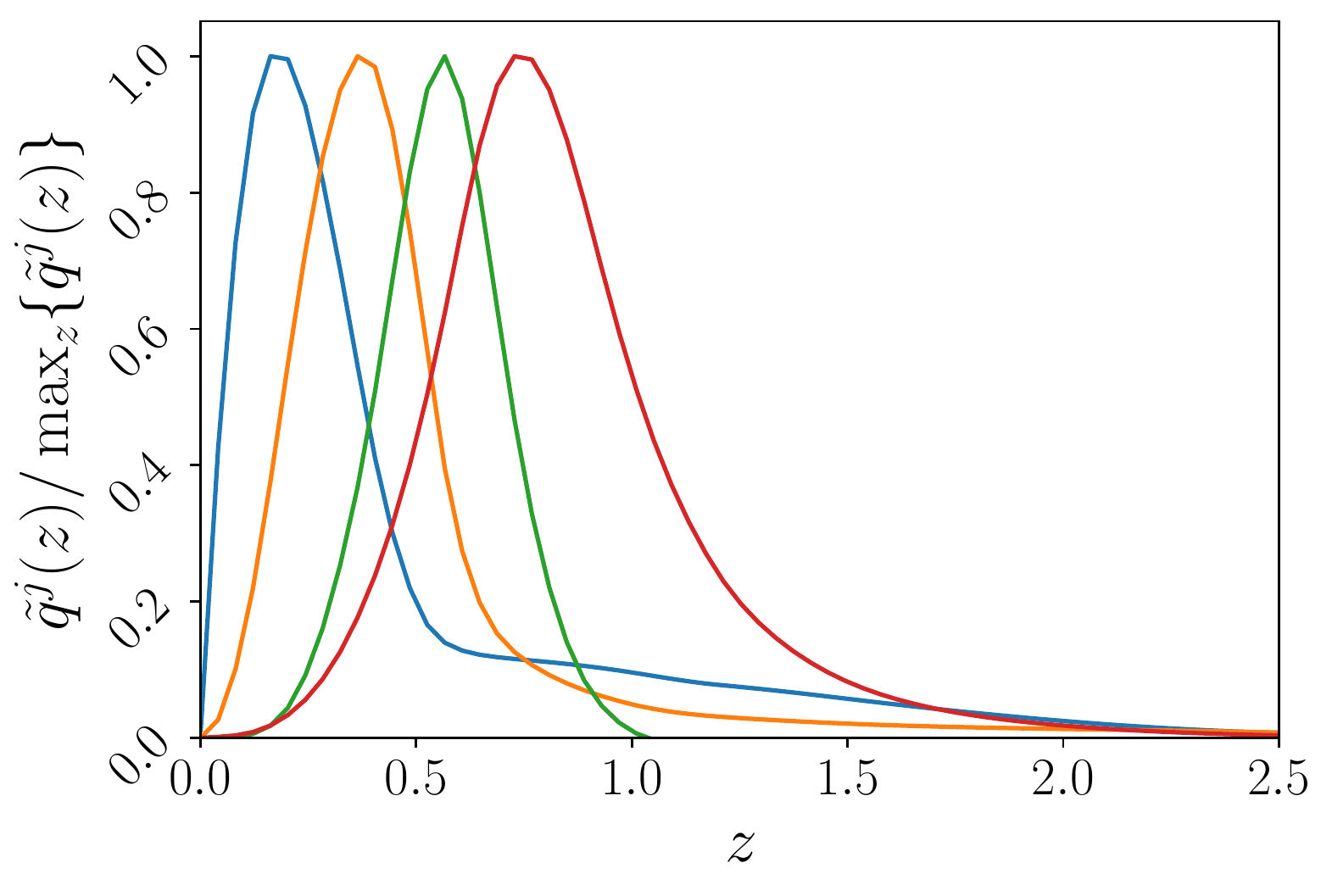}
\includegraphics[width = 5.5cm]{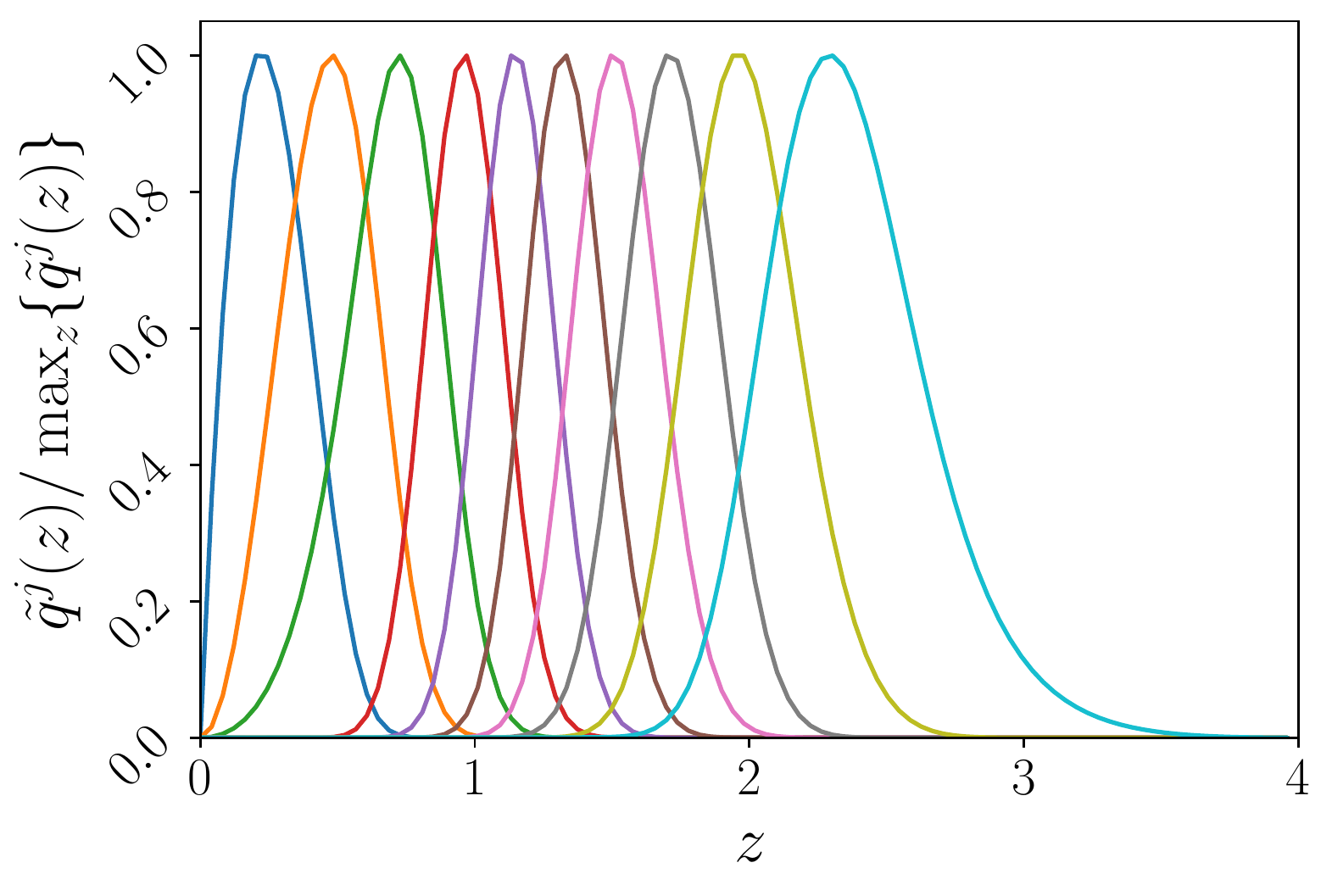}
\includegraphics[width = 5.5cm]{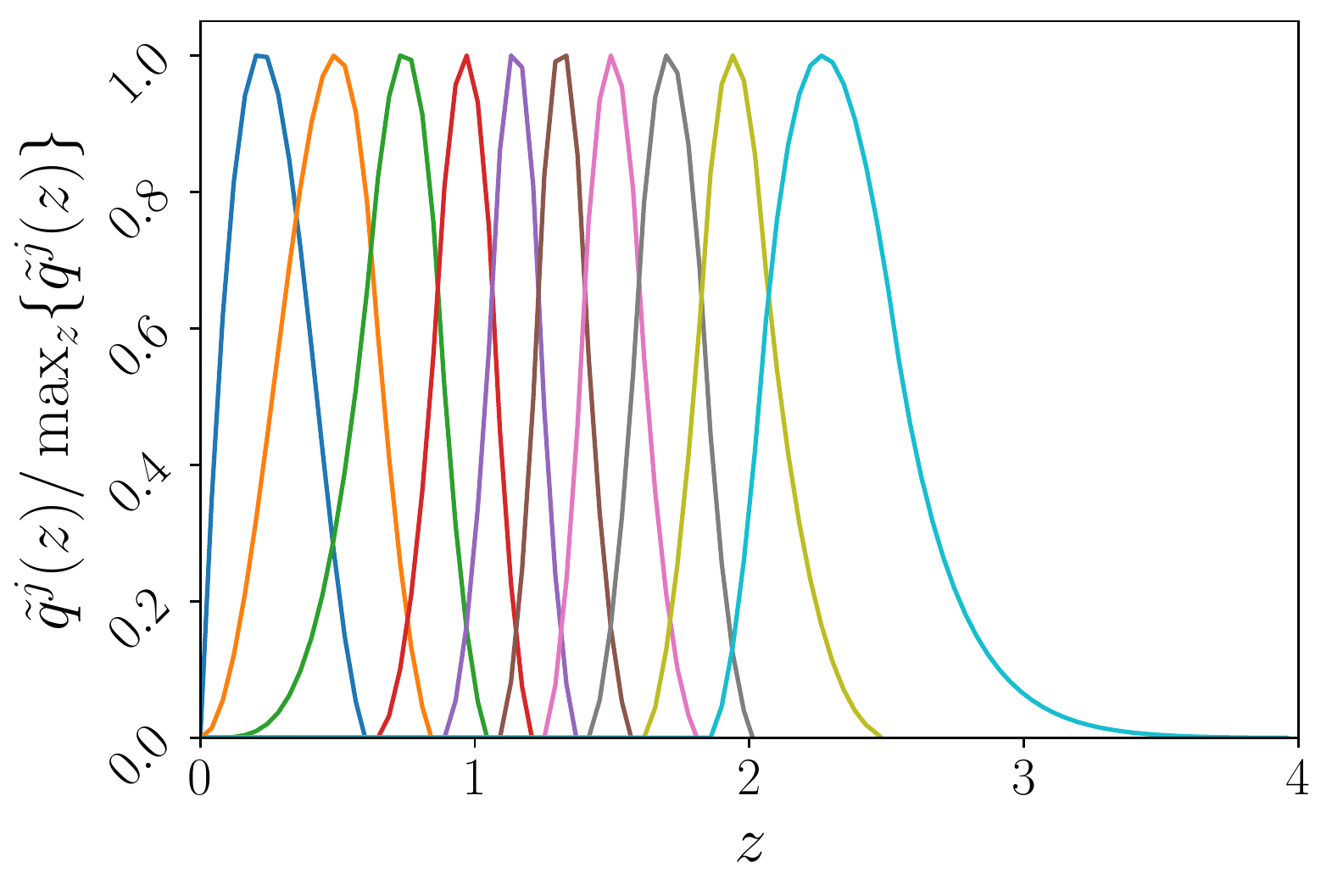}

\caption{{\bf Top row :} The radial distribution of galaxies inside tomographic bins. {\bf Middle row:} Lensing efficiency kernels, $q^i (\chi)$, for tomographic bin $i$ (the peak is normalized to 1). These are broad in redshift and there is significant overlap between bins. {\bf Bottom row:} The BNT reweighed lensing effeciency kernel, $\tilde q^i (\chi)$, (the peak is normalized to 1). These are narrow in redshift compared to the unweighted case and there is less overlap between bins. For each tomographic bin this makes it possible to relate physical structure scales, $x$, to angles, $\theta$, as well as reducing correlations between bins. {\bf Left column:} The 4 tomographic bins used in the DESY1 analysis. {\bf Middle column:} 10 tomographic bins representative of a Stage IV experiment. {\bf Right column:} Same as the middle column, but with no photometric redshift error. The BNT transform works best -- in the sense that transformed kernels are narrow and have minimal overlap -- for a large number of tomographic bins with small photometric redshift error.}
\label{fig:kernels}

\end{figure*}

 \begin{figure*}[!hbt]
 \includegraphics[scale = 0.9]{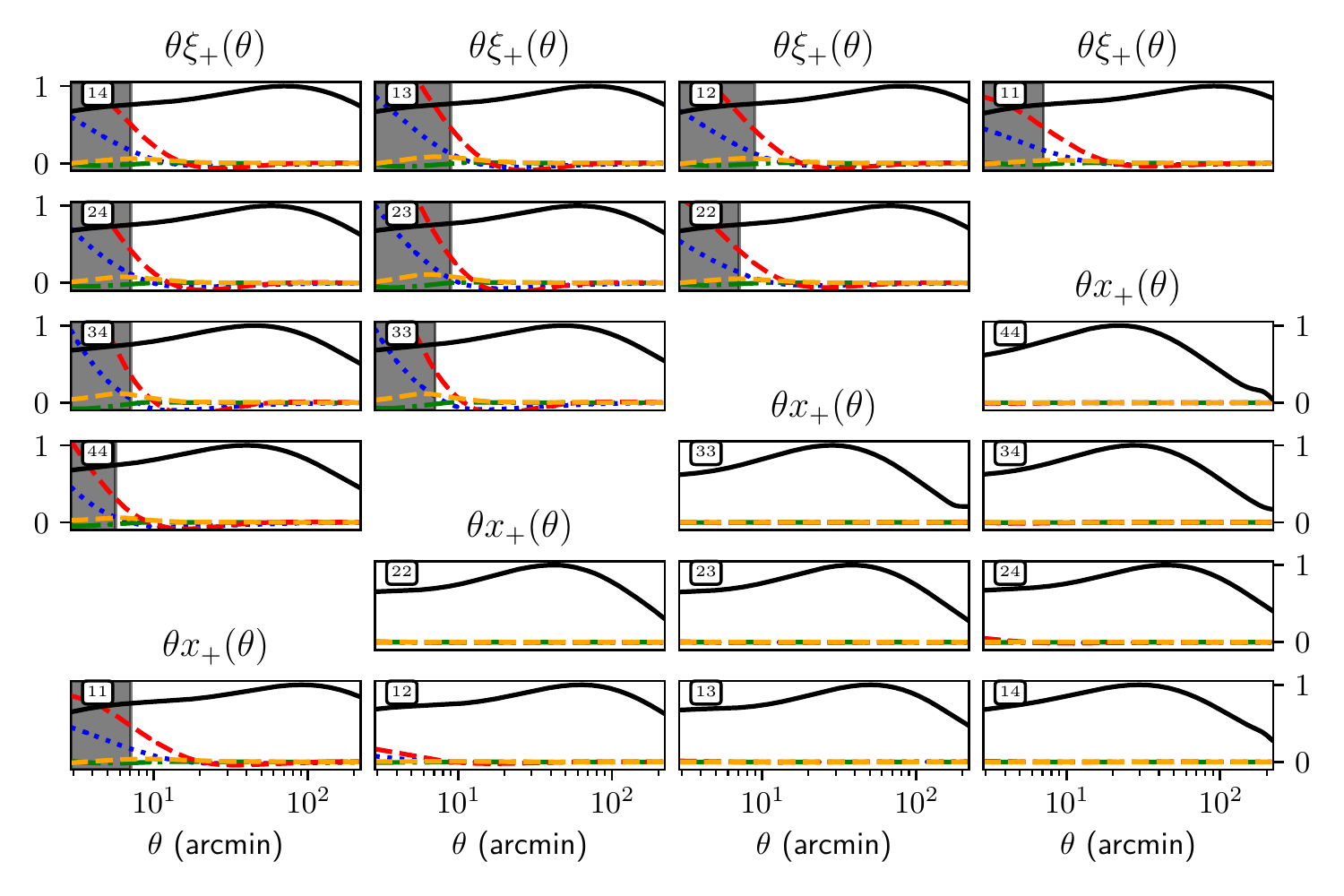}
  \includegraphics[scale = 0.9] {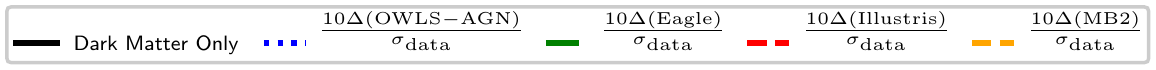}
  \includegraphics[scale = 0.9]{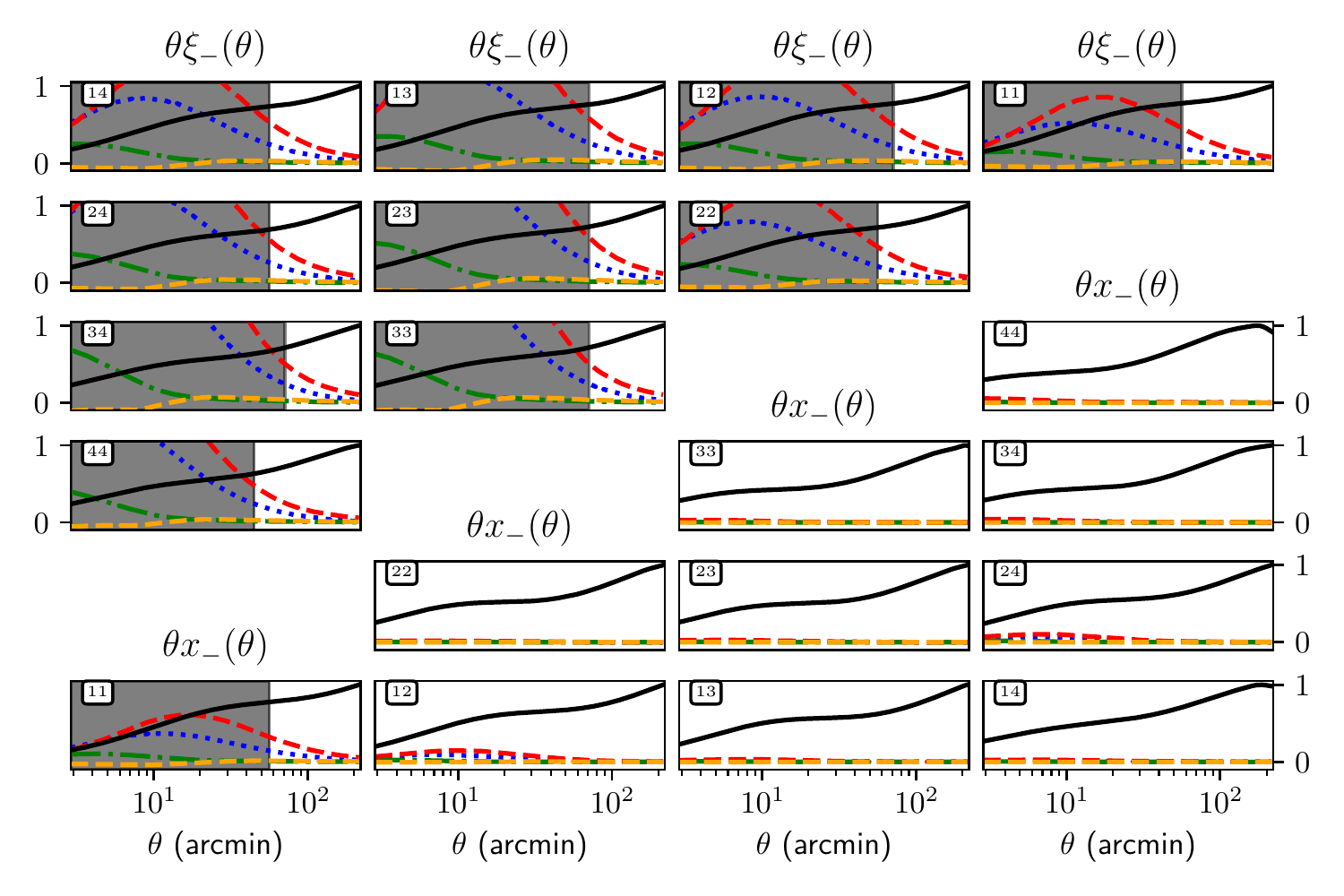}
 \caption{Solid black curves denote $\theta x_{\pm}^{ij}(\theta)$ and $\theta \xi_{\pm}^{ij} (\theta)$ (normalized so that the max value is unity) computed using the dark matter only power spectrum from {\tt Halofit}. Tomographic bin numbers are indicated in the top left hand corner of each box. The same functions are computed using power spectra from four different baryonic feedback models (see Section~\ref{sec:baryons}). For each model we plot $10 \Delta ({\rm sim}) / \sigma_{\rm data}$ where $\sigma_{\rm data}$ is the error on the data defined as the square root of the diagonal of the covariance matrix (see Section~\ref{sec:cov} for more details) and $\Delta ({\rm sim}) = d_{\rm sim} - d_{DM}$, where $d_{DM}$ is the value of the correlation function or the $x$-cut statistic computed using the dark matter-only power spectrum and $d_{\rm sim}$ is the value when baryonic feedback is included.  Grey shaded regions denote scales which are excluded from our likelihood analysis because baryonic modelling uncertainties exceed $5 \%$ of the constraining power of the survey (this is formalized in Section~\ref{sec:robustness}). Because the baryonic modelling uncertainty is concentrated at small scales in the lowest redshift bin, fewer data points must be cut when using $x_{\pm}^{ij}(\theta)$.}
 \label{fig:plus}
 \end{figure*}

 \begin{figure*}[!hbt]
 \includegraphics[scale = 0.90]{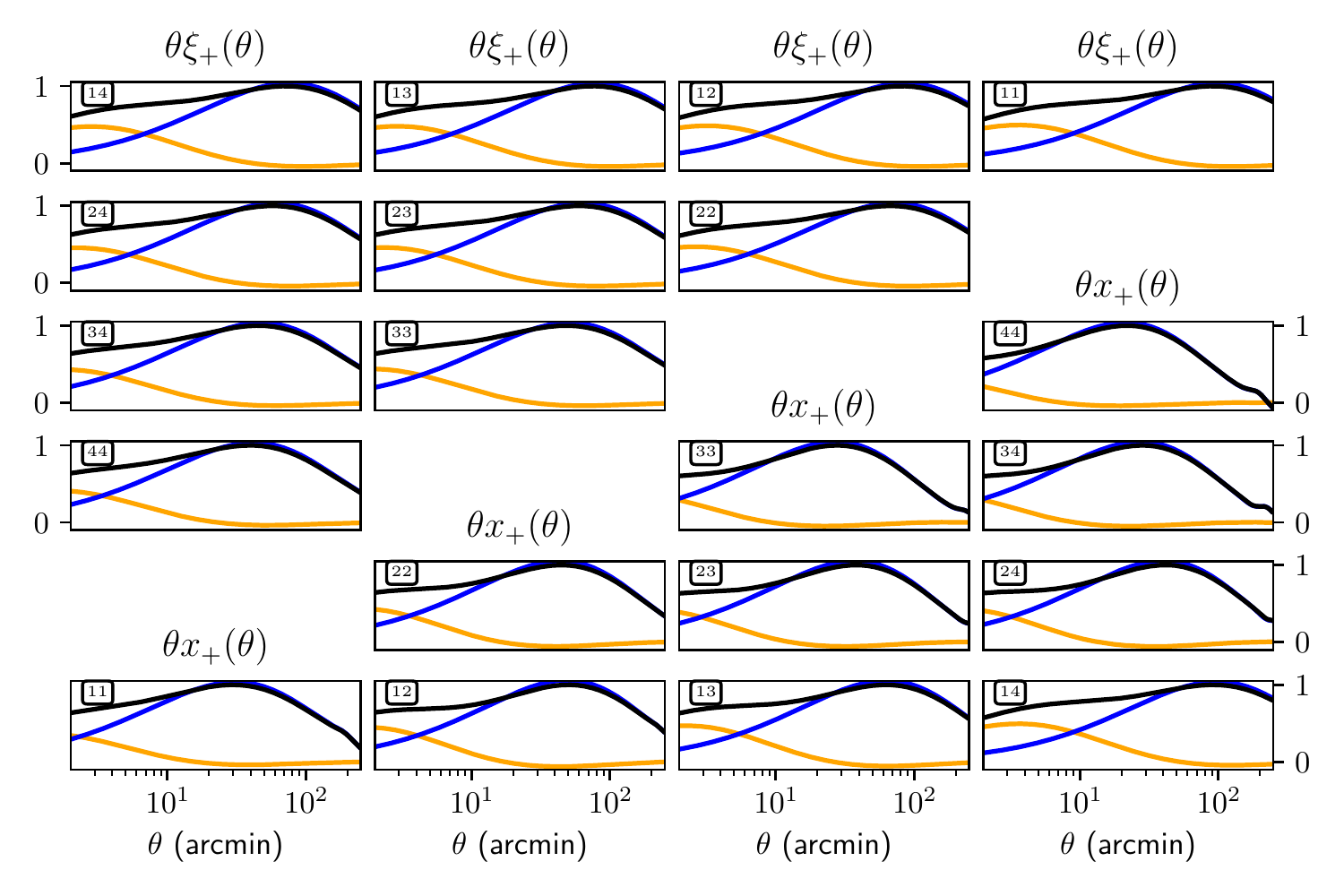}
  \includegraphics[scale = 0.90] {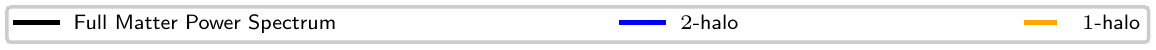}
  \includegraphics[scale = 0.90]{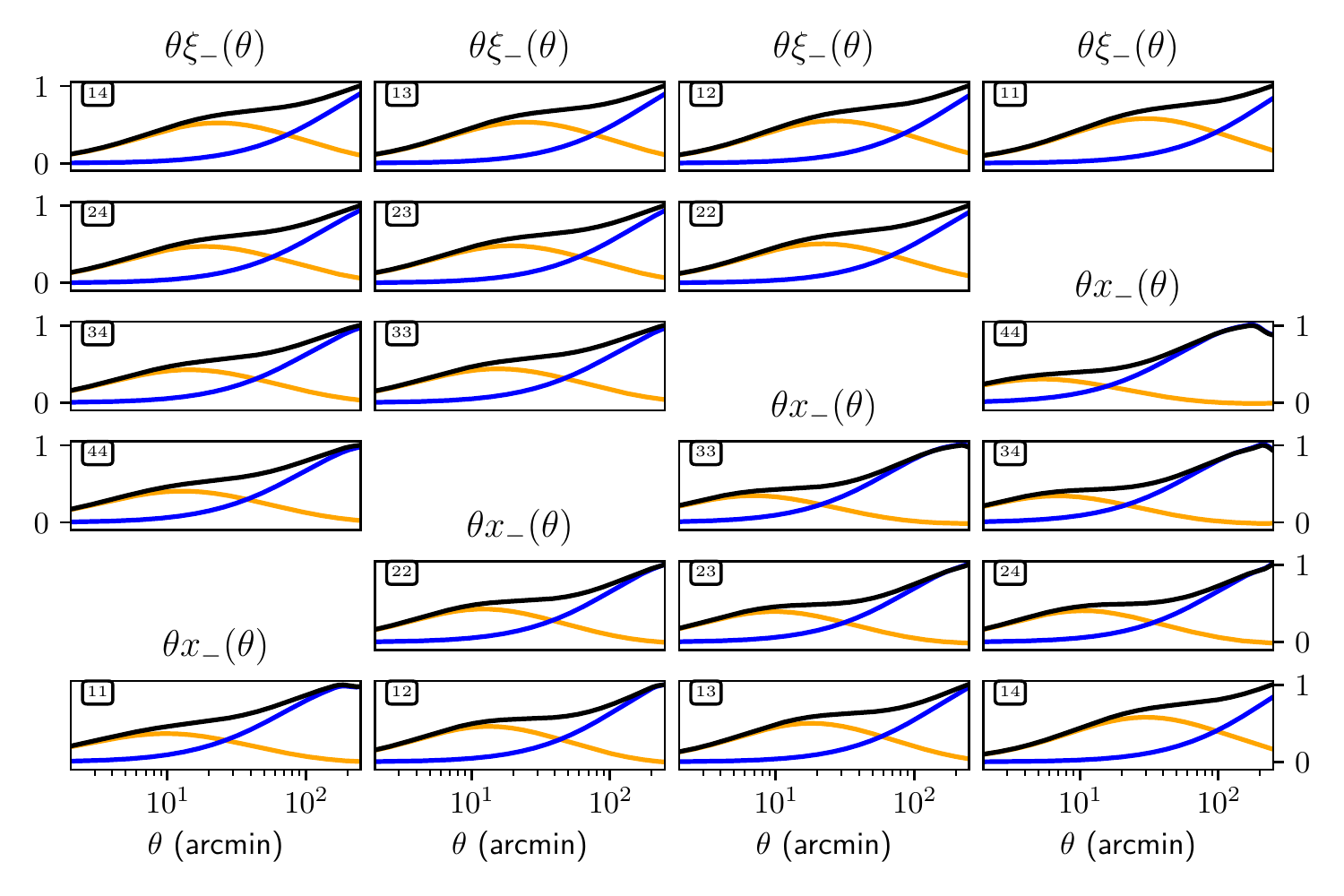}
 \caption{Solid black curves denote $\theta x_{\pm}^{ij}(\theta)$ and $\theta \xi_{\pm}^{ij} (\theta)$ (normalized so that the max value is unity) computed using the full nonlinear dark matter only power spectrum from {\tt Halofit}. The orange curve denotes the 1-halo (non-linear) matter power spectrum contributions (computed using {\tt Camb}), while the blue curve give the 2-halo (linear) contributions. For the $x$-cut statistic the $2$-halo contribution is dominant down to smaller values of $\theta$, compared to the correlation function case making it easier to separate (easy to model) linear and (hard to model) nonlinear contributions. However, at present, baryonic feedback dominates modelling uncertainty at small scales~\cite{troxel2018dark}.}
 \label{fig:halo}
 \end{figure*}

To motivate the need for a transformation of the data, we plot the lensing efficiency kernels, $q^i(\chi)$, in the middle row of Figure~\ref{fig:kernels} for three separate cases. These are:
\begin{itemize}
    \item {{\bf Left:} The four DESY1 tomographic bins considered in D18.}
     \item {{\bf Middle:} $10$ tomographic bins with an equal number of galaxies per bin drawn from:
\begin{equation}
n \left( z \right)  \propto \left( z/ z_{e} \right) ^2 \exp \left( {- \left( z / z_e \right) ^ {3/2}} \right),
\end{equation}
with $z_e = 0.9 / \sqrt{2}$, smoothed by the Gaussian kernel:
\begin{equation} \label{eq:photo error}
p \left( z | z' \right) = \frac{1}{\sqrt{2 \pi} \sigma \left(z' \right)} \exp \left ({- \frac{ \left( z - z' \right) ^2  } {2 \sigma(z')^2} } \right),
\end{equation}
to account for photometric redshift uncertainty, with $\sigma(z') =0.05 \left(  1 + z' \right)$  \cite{ilbert2006accurate}. This is representative of a typical Stage IV survey. }
\item{{\bf Right:} Same as the previous case but with perfect redshift knowledge.\footnote{This would be relevant for a kinematic weak lensing survey (see~\cite{huff2013cosmic} for more details).}}
\end{itemize}
\par What is noticeable in all three cases, is that the lensing efficiency kernels are broad in redshift and there is significant overlap between kernels. This makes it very difficult to disentangle scales using the traditional two-point estimators ($\xi_\pm$ and $C_\ell$). 
\par In particular, tomographic bins at high redshift are sensitive to structure at both high and low redshifts, so that an angular scale, $\theta$, does not correspond to a unique physical scale, $x$, in the intervening lensing structure. Additionally, since the lensing kernels have significant overlap, different tomographic bins probe the same underlying structure. This motivates the Bernardeau-Nishimichi-Taruya (BNT) transformation, which makes the lensing kernels narrower in redshift, as well as reducing overlap between kernels.

\subsection{The BNT Transformation}
The BNT transformation is a linear transformation, $M$, which makes the lensing kernels compact in redshift. The new BNT reweighted kernels, $\tilde q^i(\chi)$, are given by:
\begin{equation}
\tilde q^i(\chi) = M^{ij}q^j(\chi)
\end{equation}
For any three sequential discrete source planes $j = i-2, i-2, i$ in a spatially-flat universe, at comoving distances, $\chi^{i-2} < \chi^{i-1}< \chi^i$, it was show in~\cite{bernardeau2014cosmic} that:
\begin{equation}
\sum_{j=i-2}^{i} M^{ij} =0 
\end{equation}
and
\begin{equation}
\sum_{j=i-2}^{i} \frac{M^{ij}}{\chi ^ j} =0
\end{equation}
ensures that the lensing kernel is zero outside the range $[\chi^{i-2}, \chi^i]$.
This is generalized to the continuous case by defining:
\begin{equation}
\begin{aligned}
B^{j} =& \int_{z_\text{min}}^{z_\text{max}} dz' \ \frac{n^{j} (z')}{\chi (z')}
\end{aligned}
\end{equation}
and solving:
\begin{equation} \label{eq:BNT_sys}
\sum_{j=i-2}^{i} M^{ij} =0 
\end{equation}
and 
\begin{equation} \label{eq:BNT_sys2}
\sum_{j=i-2}^{i} M^{ij} B^j =0.
\end{equation}
This is an under-constrained system of linear equations, so we take $M^{ii} = 1$ (for all $i$). The elements of $M$ can then be computed iteratively considering sequential triplets of tomographic bins, going from lowest to highest redshift. With the choice $M^{ii} = 1$, the first tomographic bin is unchanged.
\par For the four DESY1 tomographic bins, the BNT matrix $M$, is:
\begin{equation}
M = 
\begin{pmatrix}
1 & 0 & 0 & 0 \\
-1 & 1 & 0 & 0 \\
0.85 & -1.85 & 1 & 0 \\
0 & 0.25 & -1.25 & 1 \\
\end{pmatrix}.
\end{equation}
We plot the resulting BNT lensing efficiency kernels (normalized so that the max value is 1) in the bottom left of Figure~\ref{fig:kernels}. Corresponding kernels are also shown for tomographic bins from a representative Stage IV survey with and without photometric redshift uncertainty. We see that the BNT transformation performs better -- in the sense that kernels have small overlap and are compact in $z$ -- for deep surveys, with a large number of tomgraphic bins and accurate photometry. 
\par The BNT transform is a function of the background geometry, and hence, the cosmology. However the analytic solution to equations~(\ref{eq:BNT_sys})-(\ref{eq:BNT_sys2}) are formed from ratios of $B^j$, so in practice the transformation has very little cosmological dependence as found in~\cite{bernardeau2014cosmic}. For the DESY1 tomographic bins, we individually perturb $h_0$ and $\Omega_m$ by $\pm 20 \%$ relative to the baseline cosmology, to the test the response in the BNT matrix. For $h_0$, one matrix element of $M$ changes by $2\%$, but the change in all other matrix elements is less than $1 \%$. In light of this, we fix $M$ for the remainder of this work. The result that the BNT transform is effectively independent of cosmology is survey specific and should be rechecked each time the BNT transformation is used. In any case, the transformation is applied to data and the theory vector consistently, so it will not lead to bias.

\subsection{$x$-cut Cosmic Shear}
By transforming the two lensing efficiency kernels appearing in equation~(\ref{eq:CGG}), the BNT transformation can also be applied at the level of the the two-point statistics. We note $M$ is constant so it can be `pulled through' nested integrals. Hence, the BNT transformation of the correlation functions defined in equation~(\ref{eqn:xifull}) is:
\begin{equation} \label{eqn:x1}
x^{ij}_{\pm} (\theta) = M^{ik} \xi^{kl}_{\pm} (\theta) \left( M^T \right)^{lj},
\end{equation}
where repeated indices are summed over and $T$ denotes the transpose. This new statistic can be directly related to $C^{ij}(\ell)$ using equations~(\ref{eqn:xiGG}) and~(\ref{eqn:xifull}). The intrinsic alignment terms have different kernels from the $GG$ term leading to some suboptimality in the transformation. However, IA contributions account for only $\sim 10 \%$ of the signal, so this is a small effect. 
\par We refer to $x^{ij}_{\pm} (\theta)$ as the $x$-cut statistic. Since the BNT matrix is always lower traingular with ones along the main diagonal, it can be shown with repeated application of the Laplace determinant expansion that the determinant of, $M$, is always one. Hence, the BNT transformation is invertible and from equation~(\ref{eqn:x1}) one can transform freely between $x^{ij}_{\pm} (\theta)$ and $\xi^{ij}_{\pm} (\theta)$. This implies in the absence of scale cuts, after appropriately transforming the covariance, that cosmological information is preserved. This is confirmed in Section~\ref{sec:verification}.
\par The $x$-cut statistic enables one to remove sensitivity to structure below some physical scale, $x$. To see this we note that because the lensing kernels are now narrow in $z$, we can define a typical angular diameter distance $d^i_A$ for each bin $i$. This can be the peak or mean value of the kernel, $\tilde q ^ i (\chi)$. Then, to remove sensitivity to scales below some physical scale, $x$, for tomographic bins $i,j$ we cut all angular scales, $\theta$, such that:
\begin{equation}
\theta < {\rm min} \{ x / d^i_A, x / d^j_A \}.
\end{equation}
This is the configuration space analog of $k$-cut cosmic shear~\cite{taylor2018k-cut} where we cut angular wavenumber $\ell > k r$ to remove sensitivity to small structure scales with wavenumbers greater than $k$.
\par One could choose the physical scale $x$ to be some fraction of $r_{200}$ of a `typical' cluster. However, in this paper, rather than removing sensitivity to some predefined physical scale, $x$, we instead choose to cut scales where the discrepancy between baryonic feedback models is large. This is made precise in Section~\ref{sec:robustness}.
\par In Figure~\ref{fig:plus} we plot the theoretical correlation functions, $\xi_\pm ^{ij} (\theta)$, and the $x$-cut cosmic shear statistics, $x_\pm ^{ij} (\theta)$, for the $4$ DESY1 tomographic bins. In the $x$-cut case, the majority of the information lies in the tomographic autocorrelation data points since the BNT transformation removes cross-bin correlations by construction. 
\par We generate the functions $x_\pm^{ij} (\theta)$ and $\xi_\pm^{ij} (\theta)$ (normalized so that their max value is unity) using the dark mater only matter power spectrum from {\tt Halofit}. We also compute the same functions using matter power spectra from four different baryonic feedback scenarios (see Section~\ref{sec:baryons}) and plot a measure of difference between the baryonic and dark matter only case relative to the survey error. Specifically we show $10 \Delta ({\rm sim}) / \sigma_{\rm data}$ where $\sigma_{\rm data}$ is the error on the data defined as the square root of the diagonal of the covariance matrix (see Section~\ref{sec:cov} for more details) and $\Delta (\rm sim)$ is given by:
\begin{equation}
    \Delta ({\rm sim}) = {d_{\rm sim} - d_{DM}},
\end{equation}
where $d_{DM}$ is the value of the correlation function or the $x$-cut statistic computed using the dark matter-only power spectrum and $d_{\rm sim}$ is the value when baryonic feedback is included.
\par Immediately it can be seen that in the $x$-cut case, the bias between the baryonic and dark matter becomes concentrated in the lowest redshift bin autocorrelation at small angles. Meanwhile the correlation function residuals are large at small angles for every tomographic bin pair. Because $x$-cut cosmic shear sorts information by physical scale, fewer data points must be removed allowing us to place tighter cosmological constraints, while remaining robust to baryonic modelling uncertainties. This can be seen by comparing the area of the grey shaded regions which denote scales where baryonic modelling uncertainties exceed $5 \%$ of the constraining power of the survey (this is formalized in Section~\ref{sec:robustness}). 
\par In Figure~\ref{fig:halo} we show the 1-halo (orange) and 2-halo (blue) contributions for both $\xi_\pm ^{ij} (\theta)$ and $x_\pm ^{ij} (\theta)$. For the $x$-cut statistic, the $2$-halo contribution is dominant down to smaller values of $\theta$, making it easier to separate (easy to model) linear and (hard to model) nonlinear contributions than in the correlation function case. 
\par We make our code to compute the BNT transform and an $x$-cut {\tt CosmoSIS} module available at: \hyperlink{https://github.com/pltaylor16/x-cut}{https://github.com/pltaylor16/x-cut}

\subsection{$x$-cut Galaxy-Galaxy Lensing} \label{sec:3x2}
Information from the cross-correlation between foreground galaxies and background shear can reduce the impact of systematics and tighten parameter constraints. The signal is called galaxy-galaxy lensing. The tangential shear in the angular frame between the foreground and background, $\gamma_{\mathrm t}^{ij}(\theta)$, is used as an estimator for the two-point information. It is given by:
  \begin{equation}
  \begin{aligned}
  \gamma_{\mathrm t}^{ij}(\theta) &= (1+m^j)\int
  \frac{d \ell\,\ell}{2\pi}\, J_2(\ell\theta)\,\int\!\! d\chi\!
  \frac{q_{\delta_{\mathrm{g}}}^i\!\!\left(\frac{\ell+1/2}{\chi},\chi\right)
    q^j(\chi)}{\chi^2}  \\ &\times
  P\!\left(\frac{\ell+1/2}{\chi},z(\chi)\right), 
  \end{aligned}
  \end{equation}
  where $m^j$ is the multiplicative shear bias, $J_2$ is the 2nd-order Bessel
function and
\begin{equation}
\bar{n}_{\mathrm{g}}^i = \int dz\; n_{\mathrm{g}}^i(z).
\end{equation}
The clustering kernel, $q_{\delta_{\mathrm{g}}}^i(k,\chi)$, is defined by:
\begin{equation}
q_{\delta_{\mathrm{g}}}^i(k,\chi) = b^i\left(k,z(\chi)\right)\frac{n_{\mathrm{g}}^i(z(\chi)) }{\bar{n}_{\mathrm{g}}^i}\frac{dz}{d\chi},
\end{equation}
and $b^i(k,z(\chi))$ is the galaxy bias. 
\par Since a lensing efficiency kernel, $q^j$, appears only once in the above expression, the appropriate BNT-reweighted tangential shear estimator, $x_t$, is:
\begin{equation}
x^{ik}_t(\theta) = \gamma_{\mathrm t}^{ij}(\theta) (M^T)^{jk}.
\end{equation}
Similar expressions for galaxy-galaxy estimators in harmonic space can be easily derived. One can then take angular scale cut as before. We focus on the weak lensing only case for the remainder of the paper.

\subsection{$x$-cut Cosmic Shear Covariances, Likelihoods and the Likelihood Sampling Method} \label{sec:likelihood resampling}

Computing a valid inverse covariance matrix, $\widehat C^{-1}$, is one of the main challenges when using any new statistic to perform parameter inference with real data. Covariances can be computed analytically, carefully accounting for the super-sample covariance (SSC) and non-gaussian (NG) terms or using forward simulations. The former approach is theoretically demanding, while the second approach is numerically expensive because a large number of simulations are needed to compute an unbiased estimate of the inverse covariance~\cite{anderson1958introduction, hartlap2007your}.
\par In the future the $x$-cut covariance could be computed directly from the matter power spectrum and survey geometry, but we now describe a method, which we refer to as the {\it likelihood sampling method}, to sidestep these usual approaches and rapidly generate a covariance matrix for the $x$-cut cosmic shear statistic. The key idea is to sample from the likelihood of the correlation functions, which is known a priori. This is, in effect, an extremely rapid way to generate mock simulations of the $x$-cut cosmic shear statistic which can then be used to compute the covariance matrix in the usual way.
\par This method makes two well motivated approximations:
\begin{itemize}
    \item {There already exists a valid estimate of the correlation functions covariance, $\widehat C_\xi$. This is true for most surveys.}
    \item{The likelihood of the correlation functions is Gaussian. It was shown that at the level of precision of a Stage IV cosmic shear experiment, assuming the likelihood of the correlation functions is Gaussian, leads to unbiased parameter constraints~\cite{lin2019non}. A similar result for the unmasked lensing power spectrum was shown in~\cite{taylor2019cosmic}.}
\end{itemize}
\par Under these assumptions, the steps of the {\it likelihood sampling method} are as follows:
 \begin{itemize}
    \item {Generate mock realisations of $\xi_{\pm} (\theta) - \langle \xi_{\pm} (\theta) \rangle$ by drawing samples from the normal distribution with mean zero and covariance $\widehat C_\xi$, so that:
    \begin{equation}
    \Big(\xi_{\pm} (\theta) - \big \langle \xi_{\pm} (\theta) \big \rangle \Big) \sim \mathcal{N} (0, \widehat C_\xi).
    \end{equation} 
    }
    \item{Apply the BNT transformation to the samples:
    \begin{equation}
    \begin{aligned}
    \Big[\xi_{\pm} (\theta) - \big \langle \xi_{\pm} (\theta) \big \rangle \Big] &\rightarrow M \Big(\xi_{\pm} (\theta) - \big \langle \xi_{\pm} (\theta) \big \rangle \Big) M ^T \\ &= \Big[x_{\pm} (\theta) - \big \langle x_{\pm} (\theta) \big \rangle \Big]
    \end{aligned}
    \end{equation}
    to generate mock realizations of the $x$-cut cosmic shear statistic relative to its mean. 
    }
    \item{Then an estimate of the $x$-cut cosmic shear covariance, $\widehat C_x$, is given by:
    \begin{equation}
    \widehat C_x =  \Bigg \langle \Big[ x_{\pm} (\theta) - \big \langle x_{\pm} (\theta) \big \rangle \Big] \Big[ x_{\pm} (\theta) - \big \langle x_{\pm} (\theta)\big  \rangle \Big] ^T \Bigg \rangle
    \end{equation}
    
    }
\end{itemize}
A similar method can be used to generate a $k$-cut cosmic shear covariance matrix.
\par Since the $x$-cut cosmic shear statistic is just a set of linear transformation of the correlation functions (potentially with cut scales), its likelihood, $\mathcal{L}_x$ is also Gaussian and can be written:
\begin{equation} \label{eq:gauss}
\text{ln } \mathcal{L}_x\left(p \right) = - \frac{1}{2} \sum_{a,b} \big[ D_a - T_a\left( p \right) \big] \widehat C_{x,ab}^{-1} \big[ D_b - T_b \left( p\right) \big],
\end{equation}
where $ D_a$ is the data vector composed of the observed $\hat{x}_{\pm}^{ij}$ (see eqn (\ref{eq:estimator_xi})), $T_a\left( p \right)$ is formed from the theoretical prediction of $x_{\pm}^{ij}$ given parameters, $p$, and $\widehat C_{x,ab}^{-1}$ is the inverse of the covariance matrix. Sampling from this likelihood using common codes, such as {\tt EMCEE}~\cite{foreman2013emcee} or {\tt Multinest}~\cite{feroz2013importance}, allows us to infer cosmological parameters.

\section{DES Year 1 Data and Covariance} \label{sec:DESY1}

\subsection{Shape Catalog}
We use the DESY1 {\tt METACALIBRATION} shear catalogs~\cite{zuntz2018dark}, applying the same selection cuts as in D18. Shapes were initially measured using {\tt NGMIX}\footnote{\hyperlink{https://github.com/esheldon/ngmix}{https://github.com/esheldon/ngmix}} before being self-calibrated with {\tt METACALIBRATION}.  This catalog contains approximately $26$ million galaxies over $1321 \ \text{deg}^2$ with a number density $5.5$ galaxies per $\text{arcmin}^2$~\cite{troxel2018dark}. 
\par {\tt METACALIBRATION} works by taking a noisy shear estimator, applying artificial shears and remeasuring the shape~\cite{sheldon2017practical,huff2017metacalibration}. This gives the shear response, $R$, which is used to find unbiased estimates of two-point correlation functions~\cite{sheldon2017practical}. While {\tt METACALIBRATION} can deal with most sources of bias, blending of galaxy induces a multiplicative bias~\cite{zuntz2018dark}. We account for this by taking the same prior on the multiplicative biases, $m^j$, suggested in~\cite{zuntz2018dark} and used in D18.

\subsection{Photometric Redshift Catalog}
We use the photometric redshift estimates and priors on the redshift biases, $\Delta z ^i$, found in~\cite{hoyle2018dark}. Photometric redshifts (photo-zs) were estimated using the Bayesian Photometric Redshift (BPZ) code  calibrated on high photometric-resolution images from the 30-band COSMOS survey field~\cite{laigle2016cosmos2015}. 
\par It is worth noting that~\cite{hildebrandt2020kids+, joudaki2019kids+} find that calibrating directly on spectroscopic data, rather than the high-photometric resolution COSMOS fields leads to lower estimates of $S_8$ and hence a larger tension with the Planck measurements in the $S_8 - \Omega_m$ plane. We choose to use the COSMOS-calibrated photo-zs to maintain consistency with D18.

\subsection{Data Vector}
\par The shear two-point correlation functions are defined as the sum (or difference) of the tangential, $\gamma_t$, and perpendicular, $\gamma_\times$, shear autocorrelations:
\begin{equation}
\xi^{ij}_\pm = \langle\gamma^i_t\gamma^j_t\rangle \pm
\langle\gamma^i_\times\gamma^j_\times\rangle .
\end{equation}
In D18, the correlation functions were estimated from the catalog by:
\begin{equation}
\hat{\xi}^{ij}_{\pm}(\theta) = \frac{\sum_{ab}  \left[ \hat{e}^i_{a,t}(\theta) \hat{e}^j_{b,t}(\theta) \pm \hat{e}^i_{a,\times}(\theta) \hat{e}^j_{b,\times}(\theta) \right]}{\langle R^i \rangle \langle R^j \rangle},
\label{eq:estimator_xi}
\end{equation}
where $\langle R^i \rangle$ is the average response over the bin (see~\cite{sheldon2017practical} for more details) and the sum is over all pairs of galaxies $a,b$. In practice we use the publicly available DESY1 extended-scales data vector~\cite{abbott2018dark} used in the D18 analysis with $20$ bins between $2.5$ and $250$ arcmins.
\par An estimate for the $x$-cut cosmic shear statistic, $\hat x^{ij}_{\pm}$, is found by BNT-transforming the correlation function estimator so that:
\begin{equation}
\hat x^{ij}_{\pm} (\theta) = M^{ik} \hat \xi^{kl}_{\pm} (\theta) \left( M^T \right)^{lj}.
\label{eq:estimator_x}
\end{equation}

\subsection{Covariance Matrix} \label{sec:cov} 

 \begin{figure}[!hbt]
\includegraphics[width = 7.7cm]{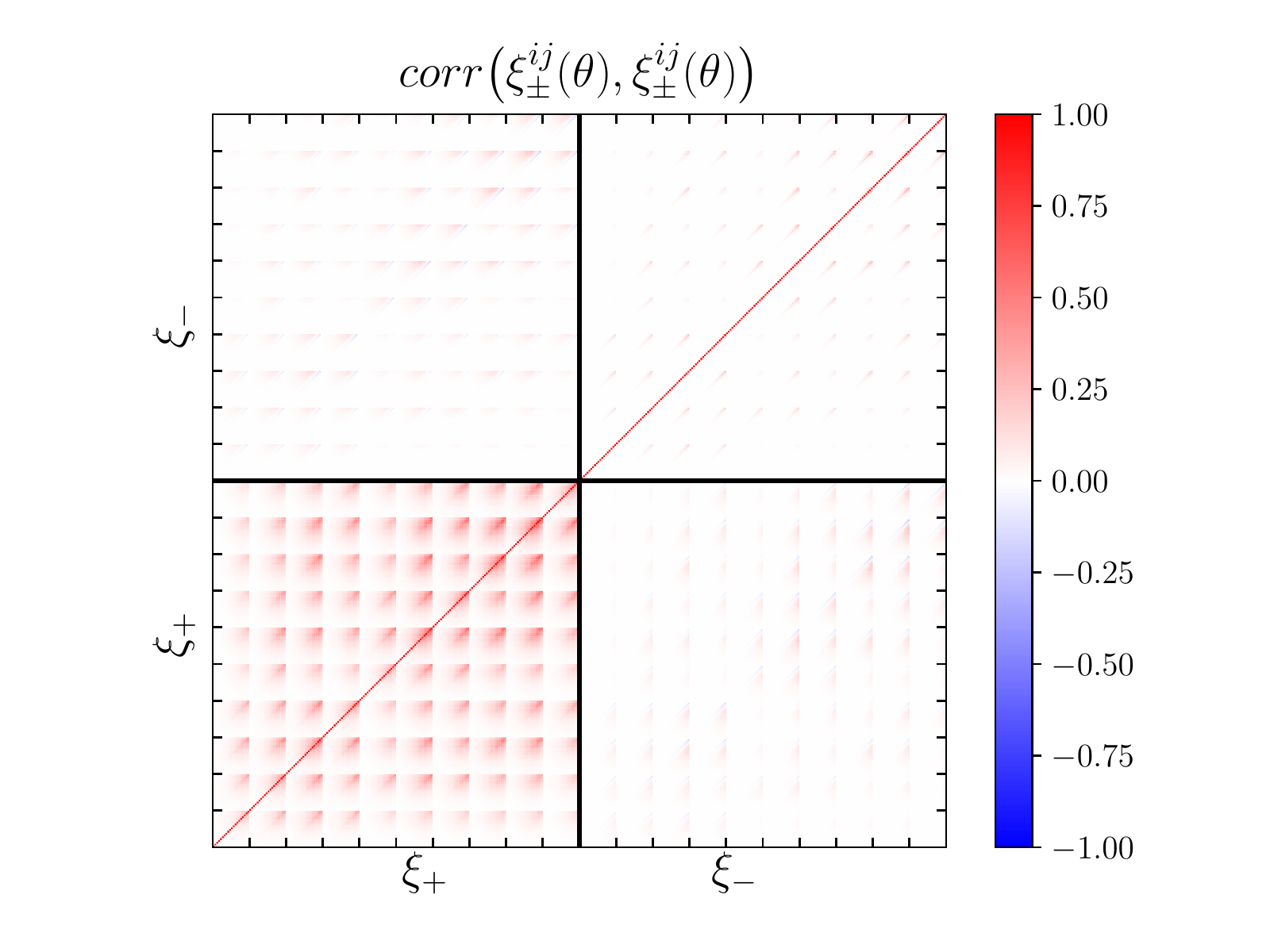}
\includegraphics[width = 7.7cm]{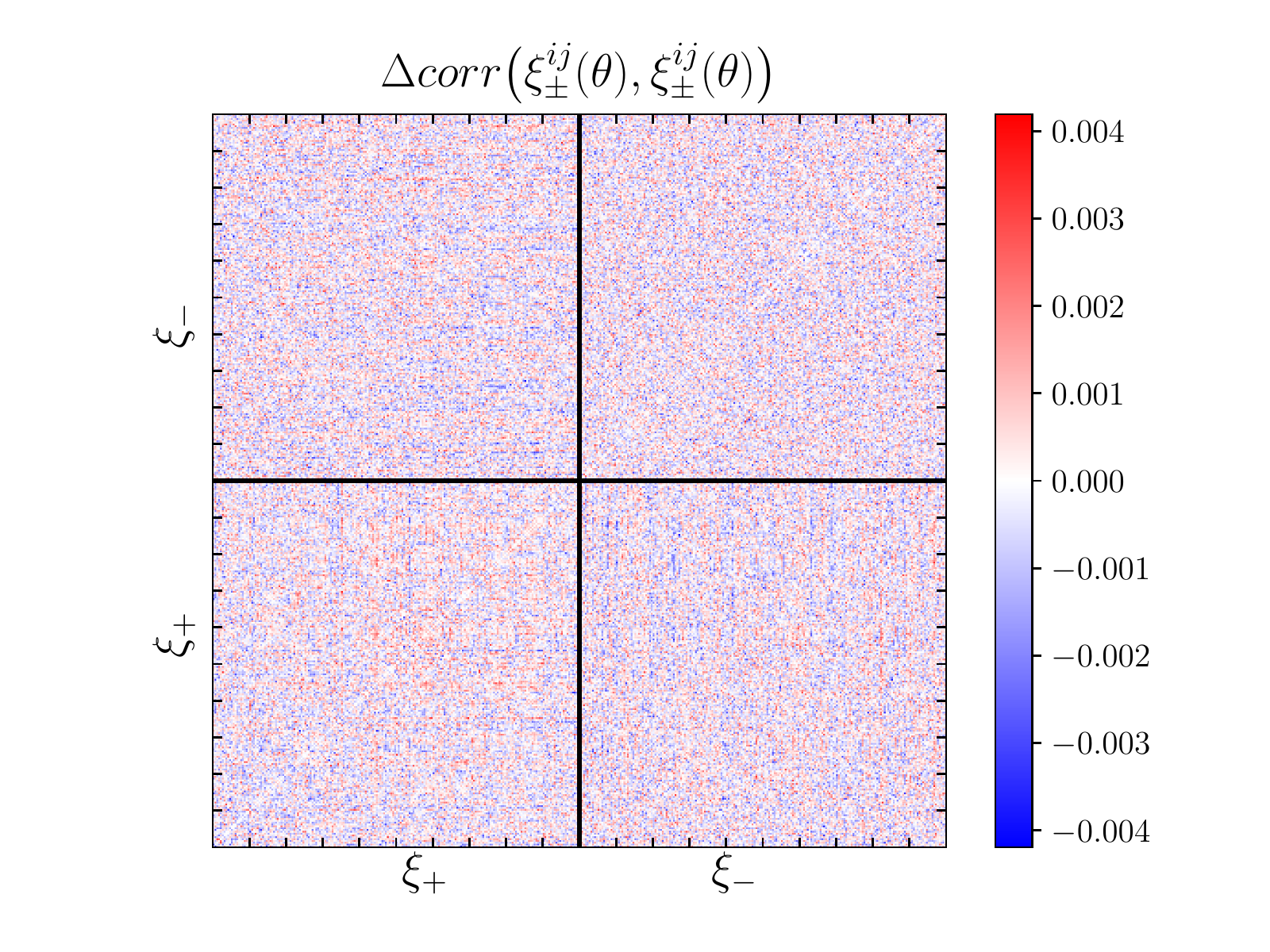}
\includegraphics[width = 7.7cm]{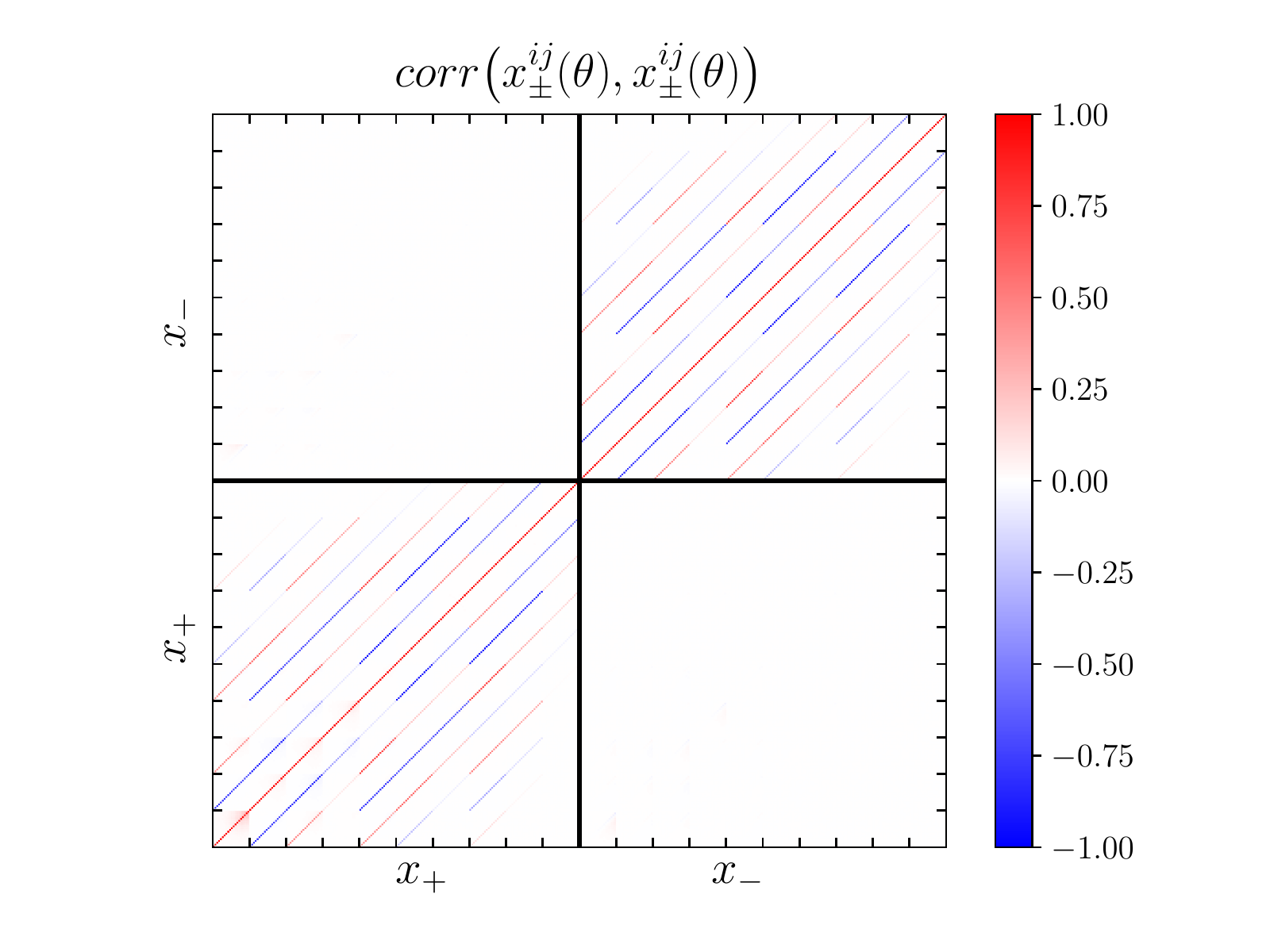}
\caption{{\bf Top:} The public DESY1 correlation matrix for $\xi^{ij}_\pm (\theta)$. The full ordering of the tomographic bin pairs, denoted by tick marks on the axes, is (11), (12), (13), (14), (22), (23), (24) etc. {\bf Middle:} The residuals between the public correlation matrix and a correlation matrix recomputed using the likelihood sampling method presented in Section~\ref{sec:cov}. Typical residuals are less than $0.5 \%$. {\bf Bottom:} An $x$-cut cosmic shear correlation matrix computed using the likelihood sampling method. The correlation matrix becomes nearly block diagonal since the BNT reweighted kernels have less overlap than before. This implies that dark matter structure at different redshifts induces the signal breaking the correlations between different tomographic bins.}
\label{fig:cov}
\end{figure}

We use the publicly available DESY1 covariance matrix used in D18. This is computed using a halo model approach and details of the calculation are given in~\cite{krause2017dark}. 
\par The corresponding correlation matrix\footnote{For covariance matrix, $C$, the correlation matrix is: $C^{ij} / \sqrt {C^{ii}} \sqrt{C^{jj}}$.} is plotted at the top of Figure~\ref{fig:cov}. The matrix is sorted into four large block matrices which give the autocorrelations, $(\xi_{+}, \xi_{+})$ and $(\xi_{-}, \xi_{-})$, for the blocks along the main diagonal and the cross-correlations $(\xi_{+}, \xi_{-})$ for the blocks off the diagonal. Inside these blocks, sub-blocks are ordered to the right (upward) by increasing redshift bin pair index $ij$ with $i \leq j$. Angular scales increase to the right (upward) within each block. 
\par To validate the likelihood sampling method presented in Section~\ref{sec:likelihood resampling}, we use it to recompute the correlation function covariance matrix. Taking $10^6$ samples from the likelihood at the fiducial cosmology with the python {\tt numpy.random.mulitivariate\_normal} function, the covariance matrix is recomputed in a few seconds. As a sanity check we ensure that the matrix is semi-positive definite by confirming the eigenvalues are all positive. The residuals between this estimate and the public $\xi$-correlation matrix are shown in middle column of Figure~\ref{fig:cov}. Typical residuals are less than $0.5 \%$. The impact of these residuals in cosmological parameter space is tested in Section~\ref{sec:verification}.
\par We also compute the $x$-cut covariance matrix directly from the DESY1 covariance using the likelihood sampling method. This takes slightly less than 3 minutes on a 2019 Macbook Pro. The computation time is dominated by applying the BNT transformation to the correlation function realizations. The corresponding correlation matrix is shown in the bottom row of Figure~\ref{fig:cov}. We note that inside each tomographic bin block, the off-diagonal correlations are small compared to the $\xi$-correlation matrix. This is because the BNT transformation has sorted scales, as intended. Given that the structure of the covariance matrix has dramatically changed, the accuracy requirements on the covariance matrix for upcoming surveys will also change (see e.g. ~\cite{taylor2013putting}). This must be investigated further in the future.

\section{Baryonic Physics} \label{sec:baryons}

\subsection{Baryonic Physics Modelling}

\begin{figure*}[!hbt]
\includegraphics[width = 8.5cm]{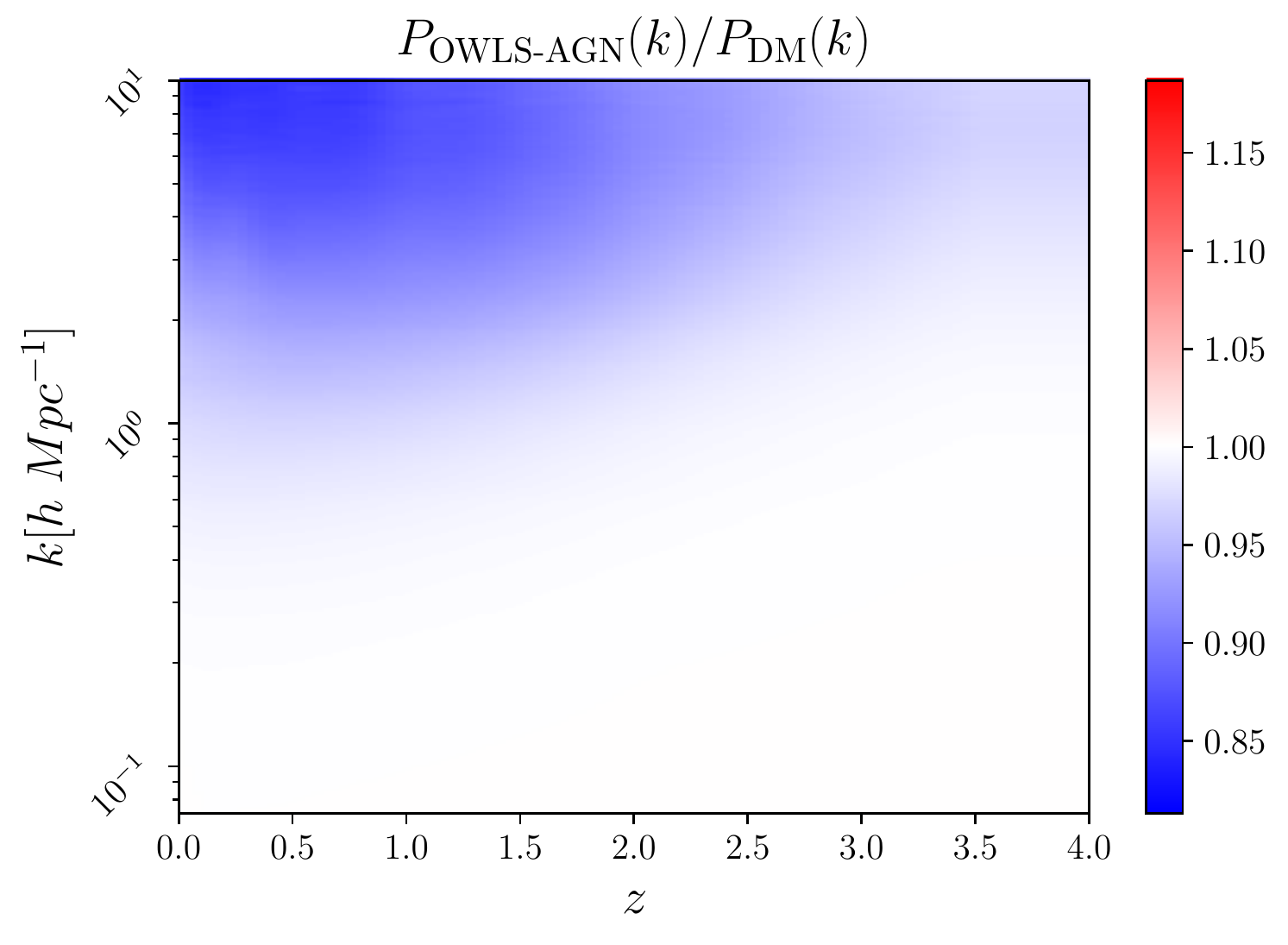}
\includegraphics[width = 8.5cm]{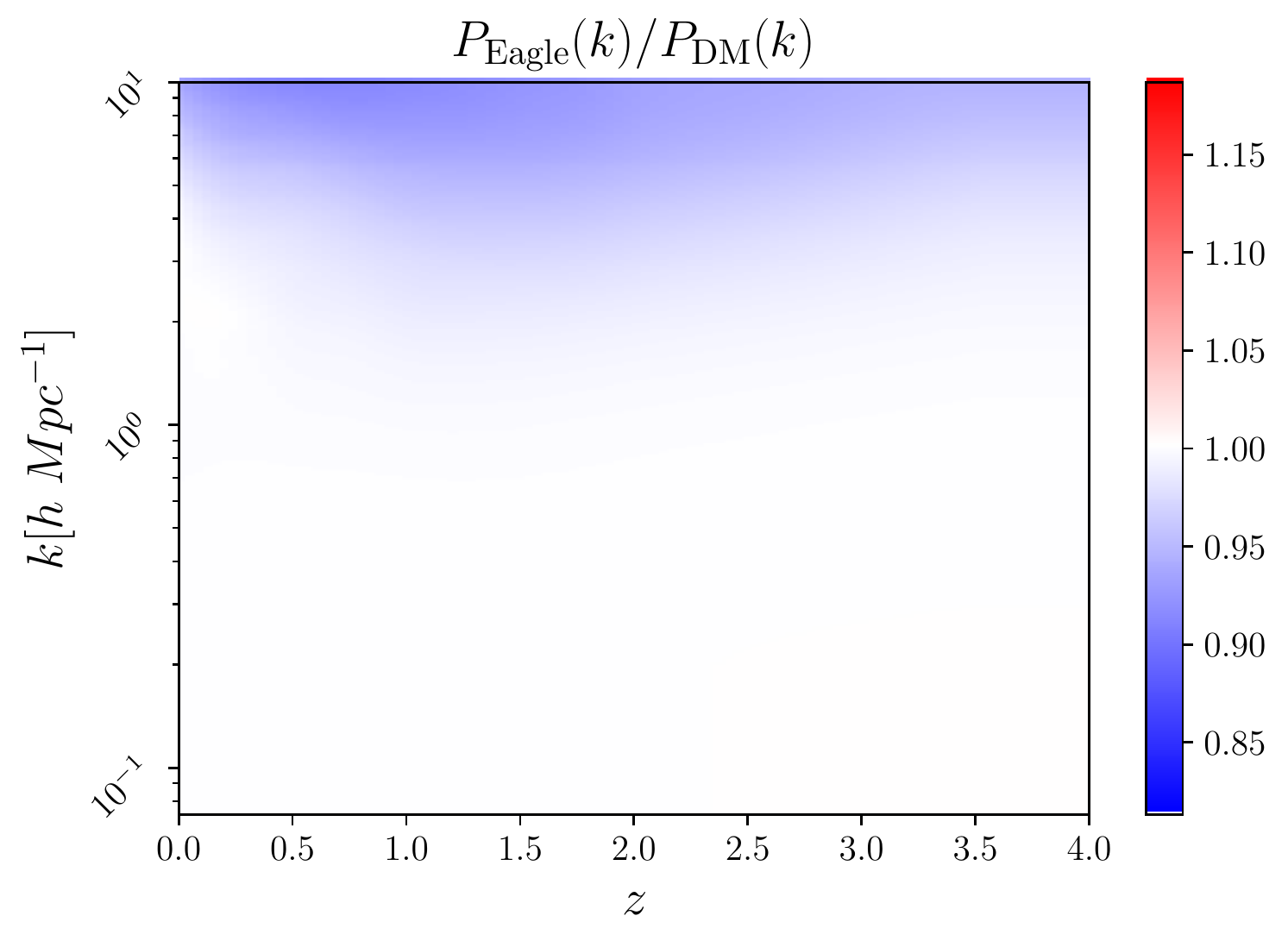}
\includegraphics[width = 8.5cm]{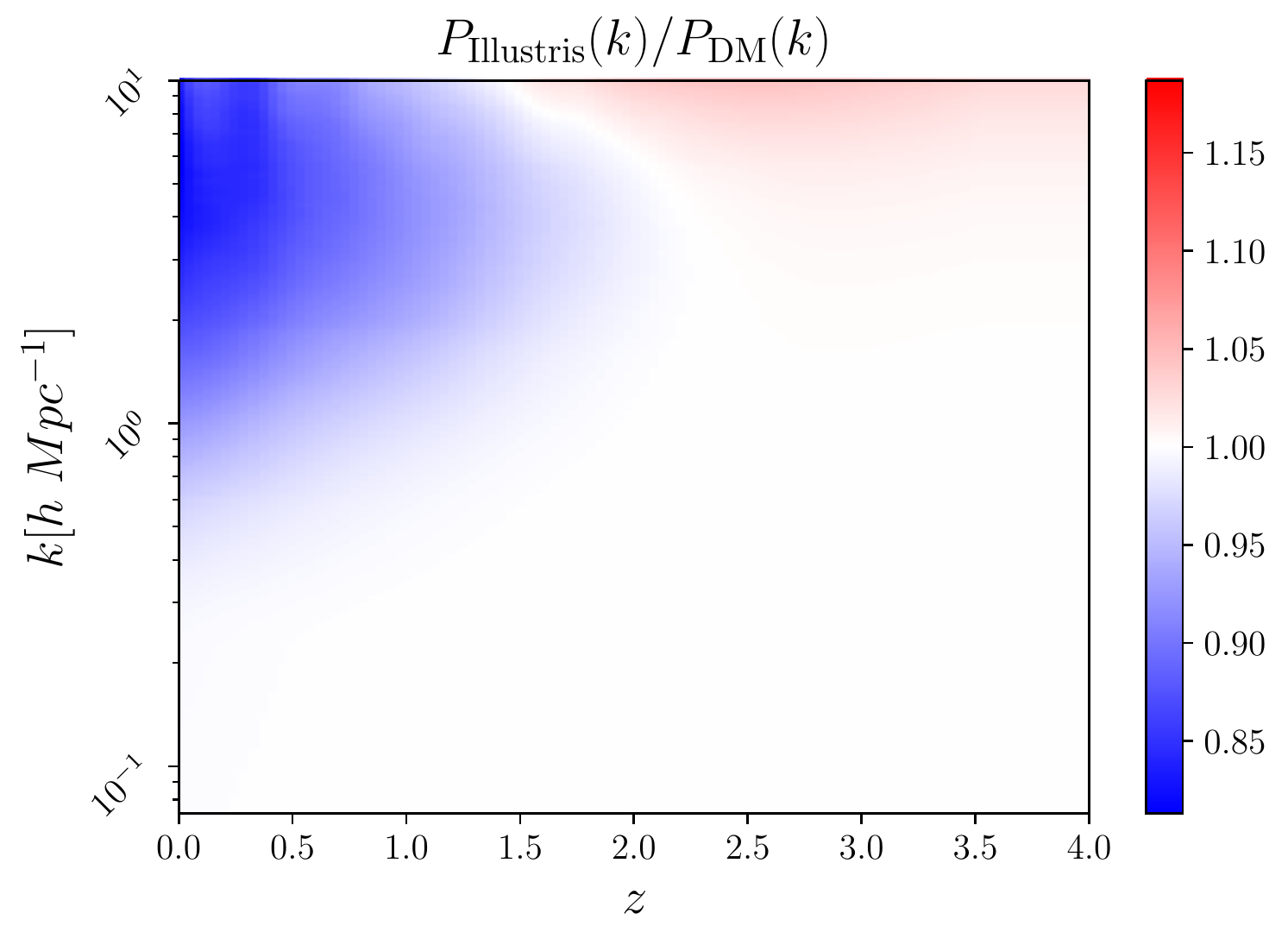}
\includegraphics[width = 8.5cm]{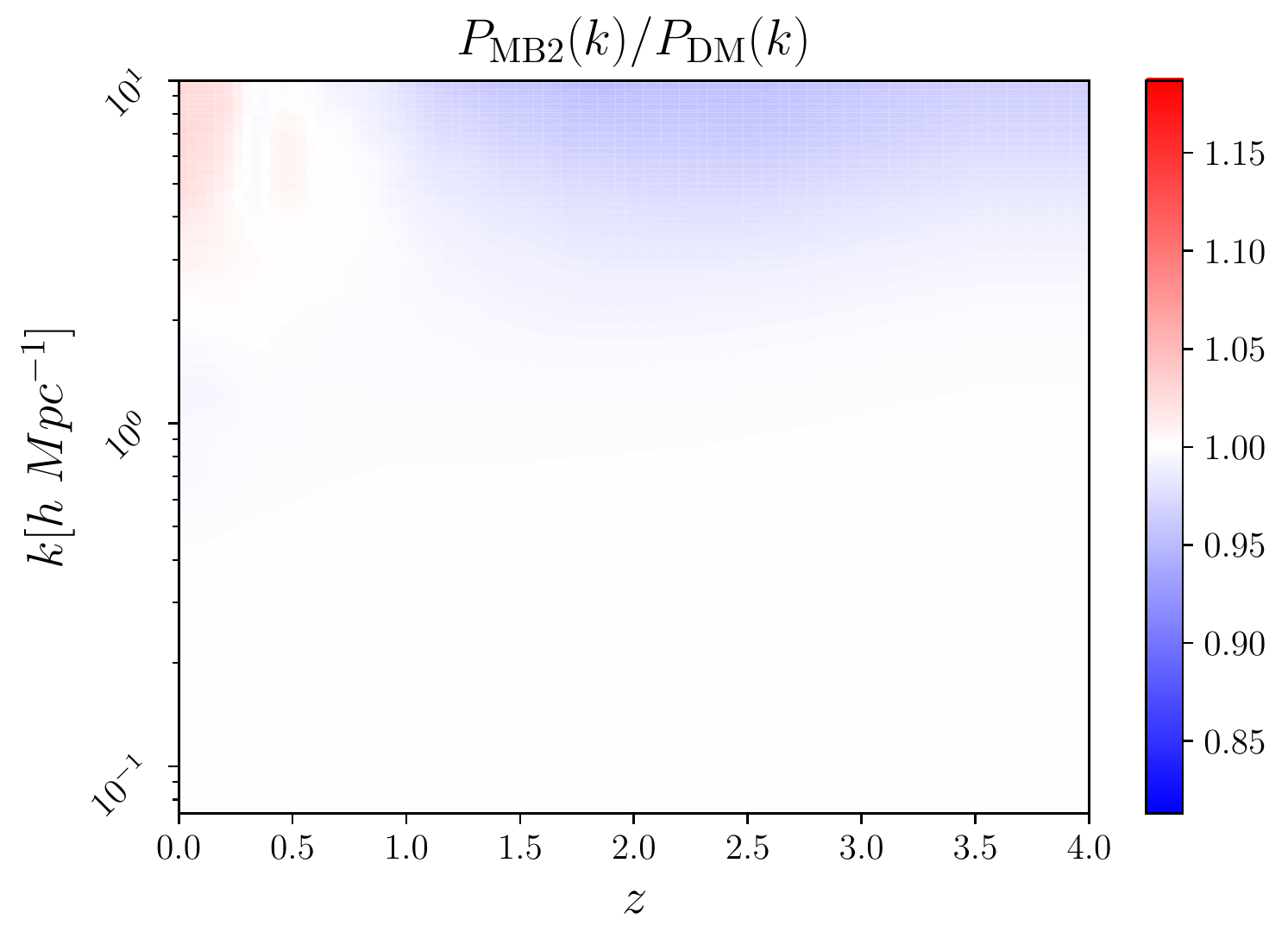}
\caption{The ratio of the time evolving baryonic matter power spectrum, relative to the dark matter only case, for the four baryonic feedback models consider in this work (see Section~\ref{sec:baryons} for more details). We make the simplifying assumption that baryonic feedback is independent of the cosmological model. This is well motivated since differences over parameter space are expected to be much smaller than between models (see~\cite{schneider2020baryonic} Figure 2). Weak lensing is primarily sensitive to scales in the range $10 ^{-1} \ h {\rm Mpc} ^{-1} \leq k \leq 10  \ h {\rm Mpc} ^{-1}$, with peak sensitivity near $1 \ h {\rm Mpc} ^{-1}$~\cite{taylor2018preparing}. Baryonic feedback suppresses power over most of this range, but in some models it is increased by the radiative cooling of gas at very small scales. There are occasional discrepancies of up to $\sim 20 \%$ between models.}
\label{fig:baryons}

\end{figure*}

 We consider four baryonic feedback models from four separate $N$-body simulations with different subgrid-physics prescriptions. These are: the OWLS simulation suite with AGN feedback~\cite{van2011effects, schaye2010physics} (OWLS-AGN), the Eagle simulation~\cite{schaye2015eagle, mcalpine2016eagle}, the Illustris simulation~\cite{vogelsberger2014introducing} and the MassiveBlack-II Simulation~\cite{khandai2015massiveblack, tenneti2015galaxy} (MB2). A detailed summary of the physical assumptions, box size and resolution of each simulation is given in~\cite{huang2018modeling}. 
 \par For cosmological parameters, $p$, we compute the temporally evolving matter power spectrum, $P(k,z;p)$, as:
 \begin{equation}
     P(k,z;p) = P_{\rm Halofit}(k,z;p) \left( \frac{P_{\rm sim }(k,z)}{P_{DM} (k,z)} \right),
 \end{equation}
 where $P_{\rm Halofit}(k,z;p)$ is the prediction from {\tt Halofit} and ${P_{\rm sim }(k,z)}/{P_{DM} (k,z)} $ is the ratio between the power spectrum from full baryonic simulations and dark matter only simulations, with matched initial conditions. We have assumed that baryonic feedback is the same at each point in cosmological parameter space. For our purposes this is a good assumption since the variation in baryonic feedback over cosmological parameter space is expected to be much smaller than between models. This can be seen by comparing Figure 2 of~\cite{schneider2020baryonic} with Figure~\ref{fig:baryons}. 
 \par Making use of the publicly available data at: \hyperlink{https://github.com/hungjinh/baryon-power-spectra}{https://github.com/hungjinh/baryon-power-spectra}~\cite{huang2018modeling}, we compute the ratios, ${P_{\rm sim }(k,z)}/{P_{DM} (k,z)} $, at different redshift slices. We use a bivariate spline to interpolate between points in $(k,z)$-space. 
 \par Using this procedure, the baryonic feedback is plotted in Figure~\ref{fig:baryons}. Weak lensing is primarily sensitive to scales in the range $10 ^{-1} \ h {\rm Mpc} ^{-1} \leq k \leq 10  \ h {\rm Mpc} ^{-1}$, with peak sensitivity near $1 \ h {\rm Mpc} ^{-1}$~\cite{taylor2018preparing}. Over most of this range, baryonic feedback suppresses structure but in some models, power is enhanced by the radiative cooling of gas at very small scales. Discrepancies between models can be as large as $\sim 20 \%$, much larger than the percent-level requirements of upcoming Stage IV surveys.

\subsection{A Note About Amplitudes: $\sigma_8$, $S_8$, $A_s$ and $A_p$} \label{sec:amplitudes}
The variance of the linear overdensity field on $ R = 8 h ^{-1} {\rm Mpc}$ scales, $\sigma_8$, is defined as:
\begin{equation}
    \sigma_8 ^2 = \int dk \ P_{\rm lin}(k) W(k,R)
\end{equation}
where $P_{\rm lin}$ is the linear power spectrum and
\begin{equation}
    W(k,R) = \frac{3k ^2}{2 \pi ^ 2 (kR)} \big[\sin (kR) - kR \cos(kR) \big].
\end{equation}
Hence $\sigma_8$ is not a physical as the lensing two-point signal is sensitive to the full nonlinear power spectrum $P(k,z)$ of the Universe including baryonic feedback corrections. Since computing the variance of the nonlinear overdensity field including baryonic corrections is a model dependent quantity, one may wish to use the spectral amplitude, $A_s$ as the physical quantity instead of $\sigma_8$ (or $S_8$). Motivated by the definition of $S_8$, we also propose a new model independent primordial amplitude parameter, $A_p$, which we define as:
\begin{equation}
A_p = A_s \left( \Omega_m / 0.3 \right) ^ 2.
\end{equation}
This breaks the degeneracy between $A_s$ and $\Omega_m$, and as we show in the remaining sections, is also constrained by cosmic shear. We plot our results in terms of $S_8$, $A_s$ and $A_p$ for the remainder of the paper.

\section{Results} \label{sec:results}

\begin{table}[hbt!]
\caption{The parameter ranges and priors used in all likelihood analyses. These choices match D18 exactly except we take a Gaussian prior on $h_0$, as in~\cite{heymans2013cfhtlens}. This ensures that the background geometry does not change significantly over parameter space so that the $x$-cut method removes sensitivity to poorly modelled scales, as intended.}
\label{table:params}
\begin{center}
\begin{ruledtabular}
\begin{tabular}{ lccccccc }
  Parameter 		&  Range & Prior   \\
  \hline

  $A_s \ (\times 10^{9})$ &  [0.5, 5.0] & Flat   \\
  $\Omega_m$ 	&  [0.1, 0.9]  & Flat  \\
  $\Omega_b$ 	& [0.03, 0.07] & Flat  \\
  $\Omega_{\nu}h^2$ &  [0.0006, 0.01] & Flat  \\
  $h_0$ 		&  [55, 90] &  $\mathcal{N} (0.7, 0.015)$  \\
  $n_s$ 		& [0.87, 1.07] & Flat  \\
  $m^1$ -- $ m^4$$ (\times10^2)$ & [$-10$, 10] & $\mathcal{N} (1.2, 2.3)$ \\
  $\Delta z^1 \ (\times10^2)$ &  [$-10$, 10] & $\mathcal{N} (0.1,1.6)$ \\
  $\Delta z^2 \ (\times10^2)$ &  [$-10$, 10] & $\mathcal{N} (-1.9,1.3)$ \\
  $\Delta z^3 \ (\times10^2)$ &  [$-10$, 10] & $\mathcal{N} (0.9,1.1)$ \\
  $\Delta z^4 \ (\times10^2)$ &  [$-10$, 10] & $\mathcal{N} (-1.8,2.2)$ \\
  $A$  &  [$-5.0$, 5.0] & Flat \\
  $\eta$ & [$-5.0$, 5.0] & Flat \\
  $z_0$  &  0.62 & Fixed \\
\end{tabular}
\end{ruledtabular}
\end{center}
\end{table}

\begin{figure*}[hbt!]
\includegraphics[scale = 0.45]{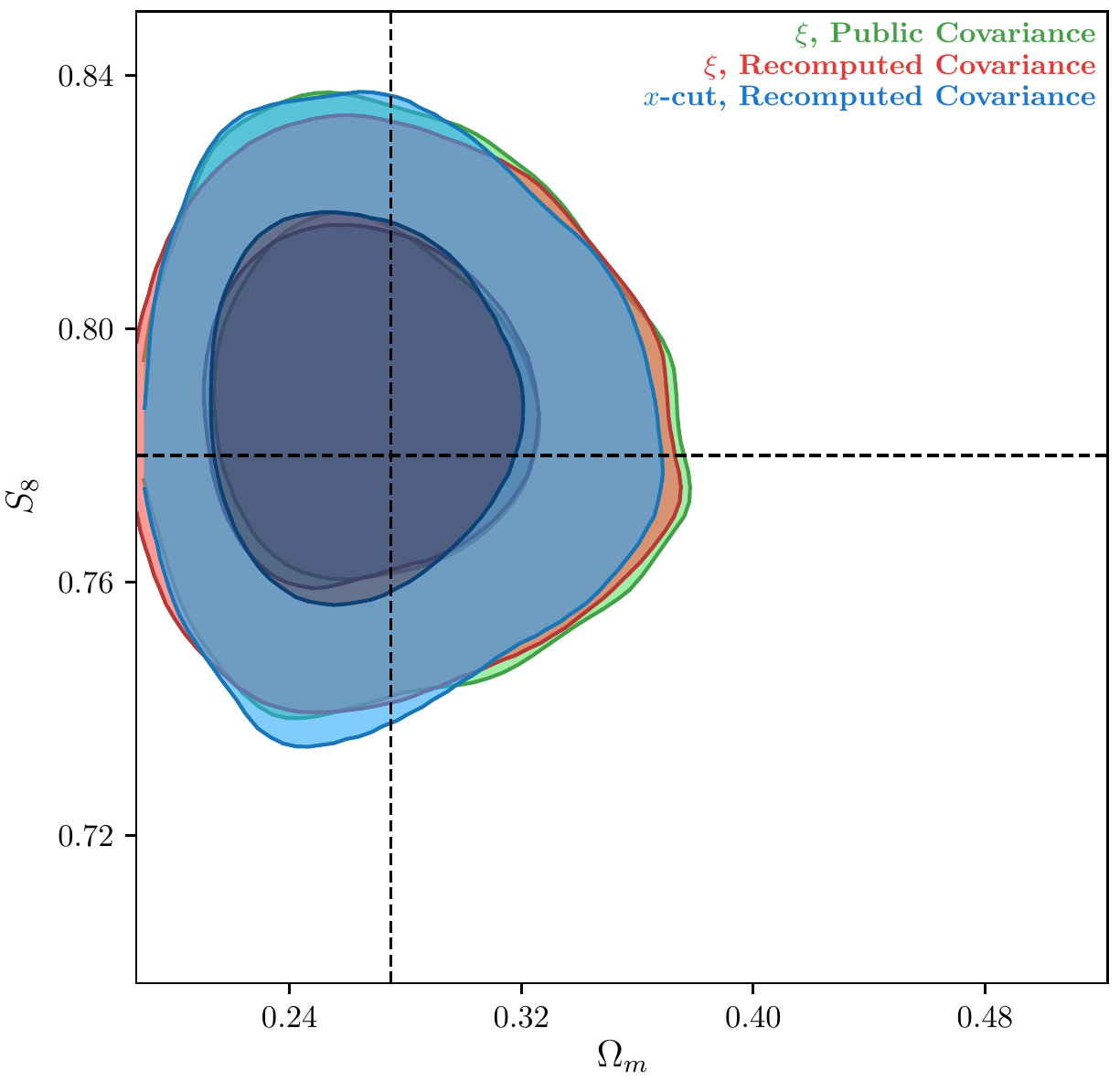}
\includegraphics[scale = 0.45]{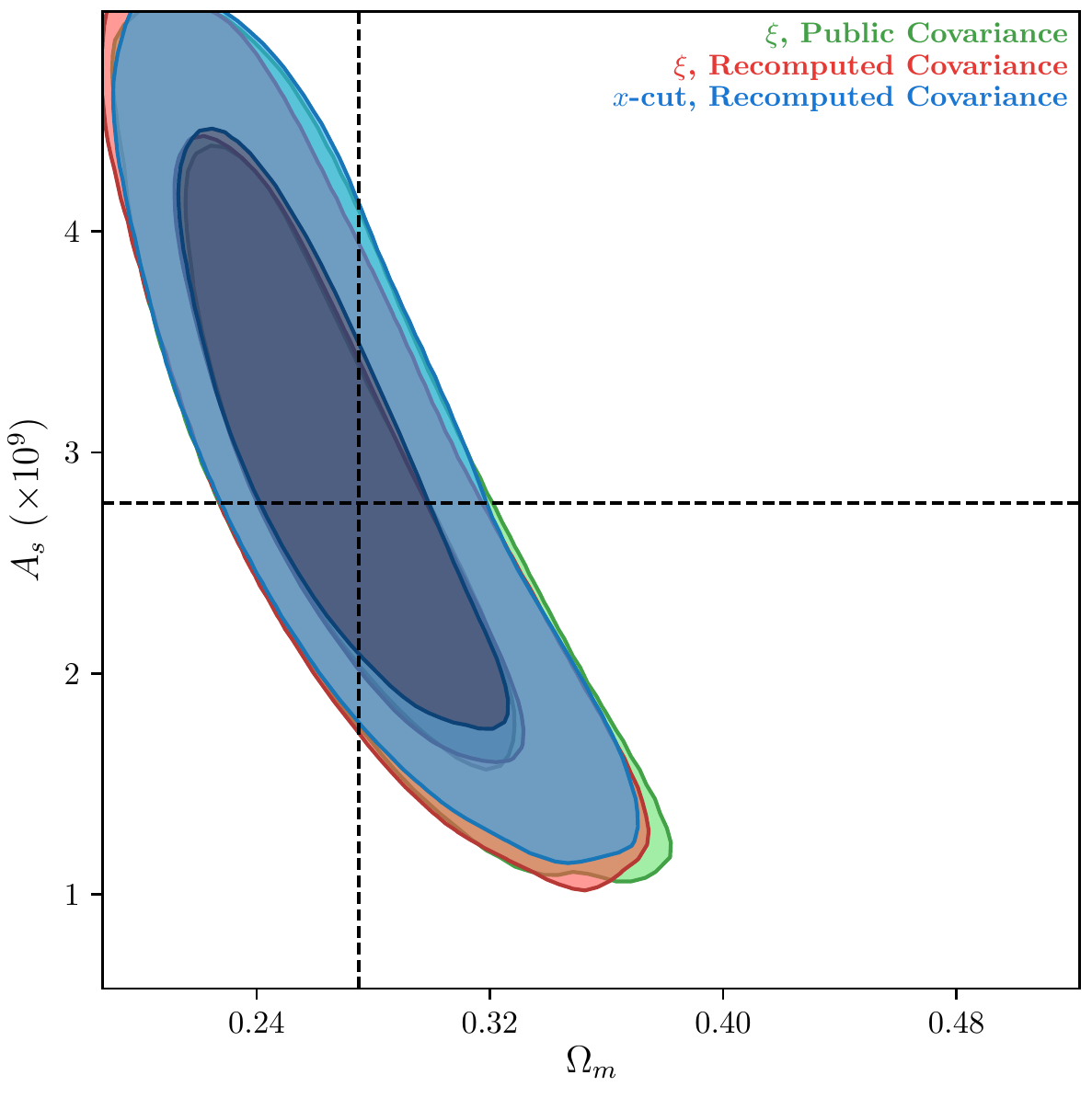}
\includegraphics[scale = 0.45]{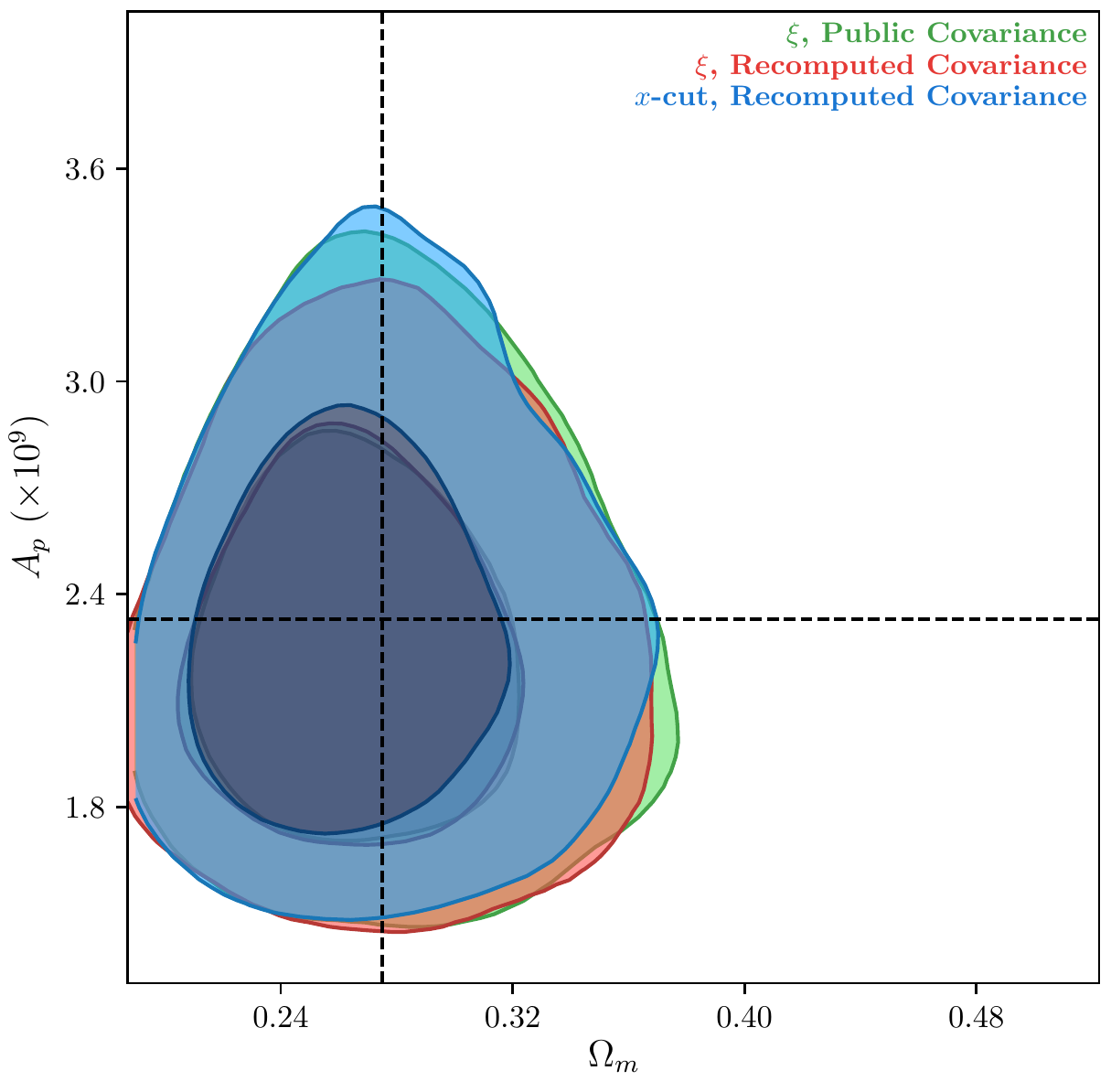}
\caption{A comparison of parameter constraints from a {\bf synthetic data realization} using the DESY1 public covariance, a $\xi$-covariance computed using the likelihood sampling method and an $x$-cut cosmic shear covariance with no scale cuts. The dotted cross indicates the input cosmology of the synthetic data. There is excellent agreement between the three likelihood chains and the input cosmology is recovered, confirming the accuracy of the computed covariances. Results are shown in the $S_8 - \Omega_m$, $A_s - \Omega_m$ and $A_p - \Omega_m$ planes. We have motivated this choice in Section~\ref{sec:amplitudes}, where $A_p$ is also defined. All contour plots are produced with {\tt ChainConsumer}~\cite{chainconsumer}.}
\label{fig:cov compare}
\end{figure*}

\begin{figure*}[hbt!]
\includegraphics[scale = 0.45]{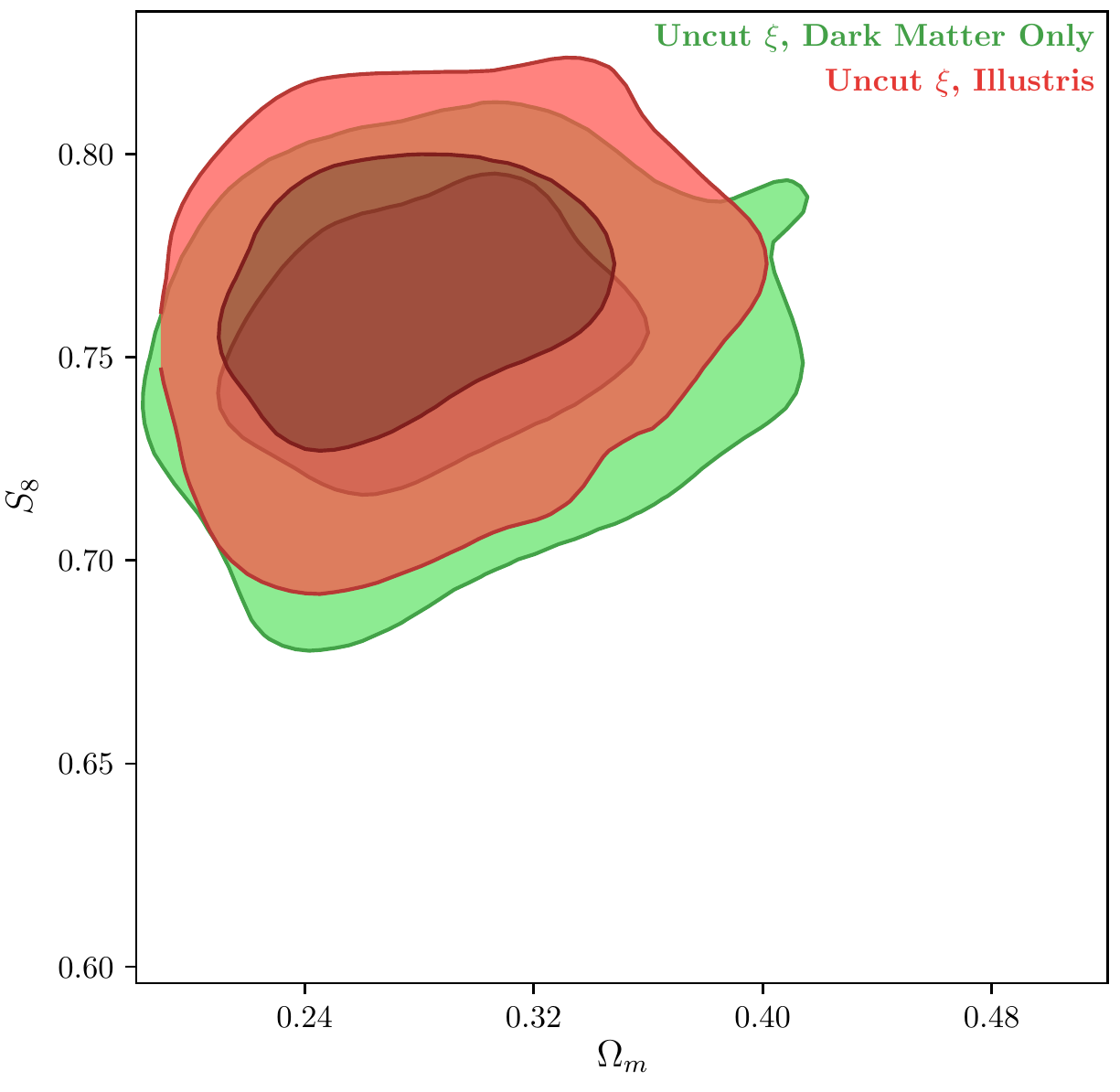}
\includegraphics[scale = 0.45]{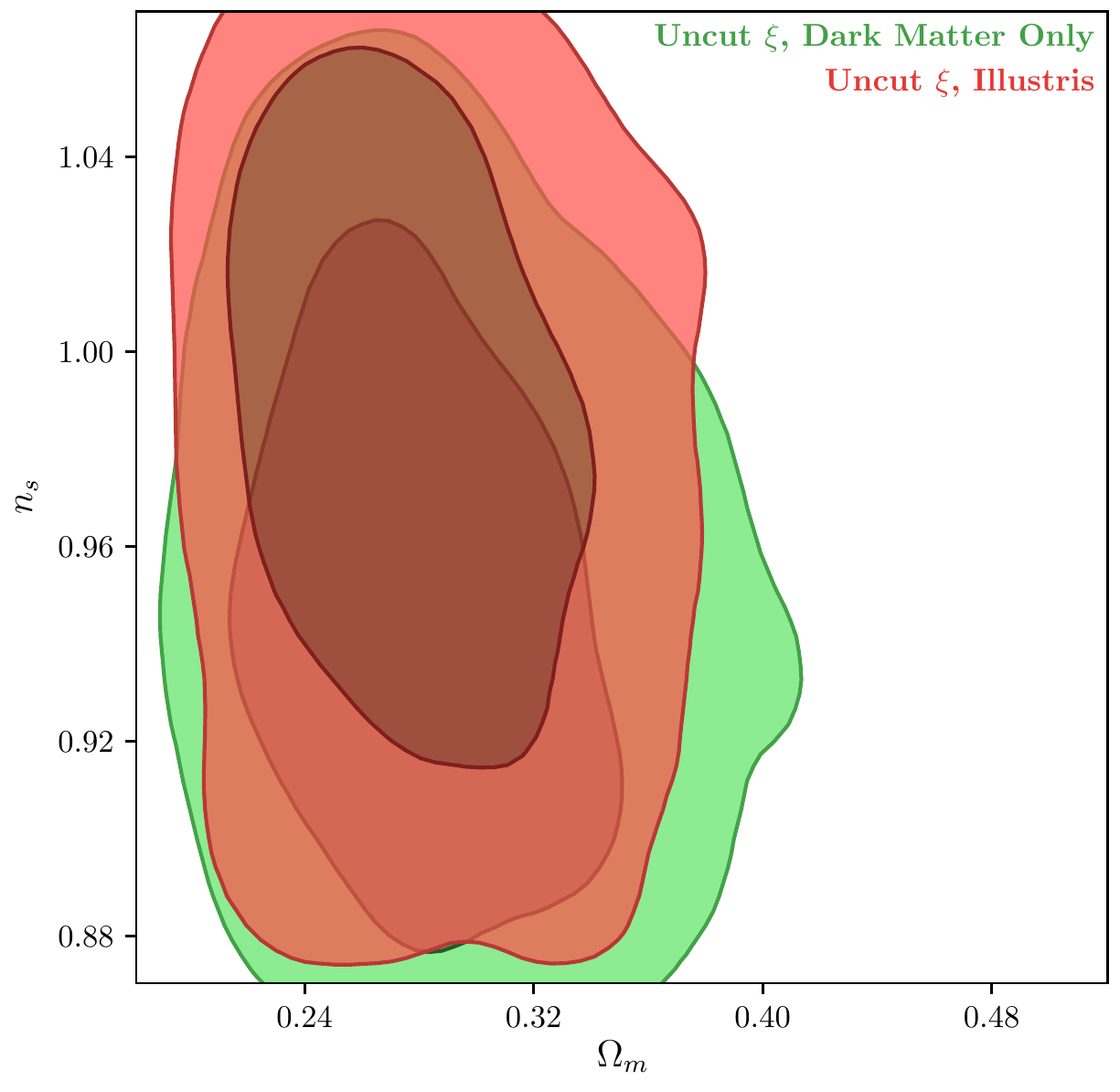}
\includegraphics[scale = 0.45]{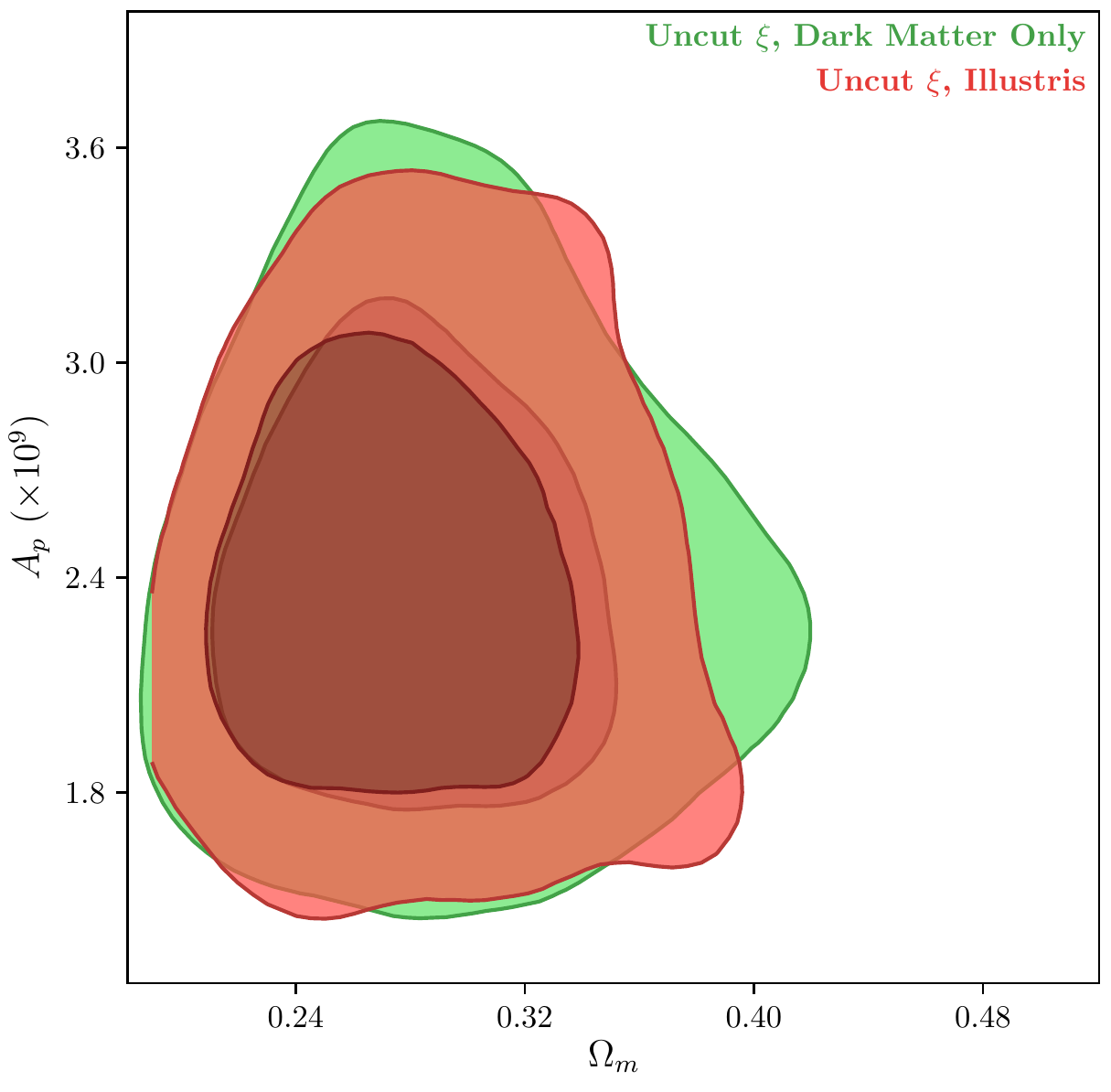}
\includegraphics[scale = 0.45]{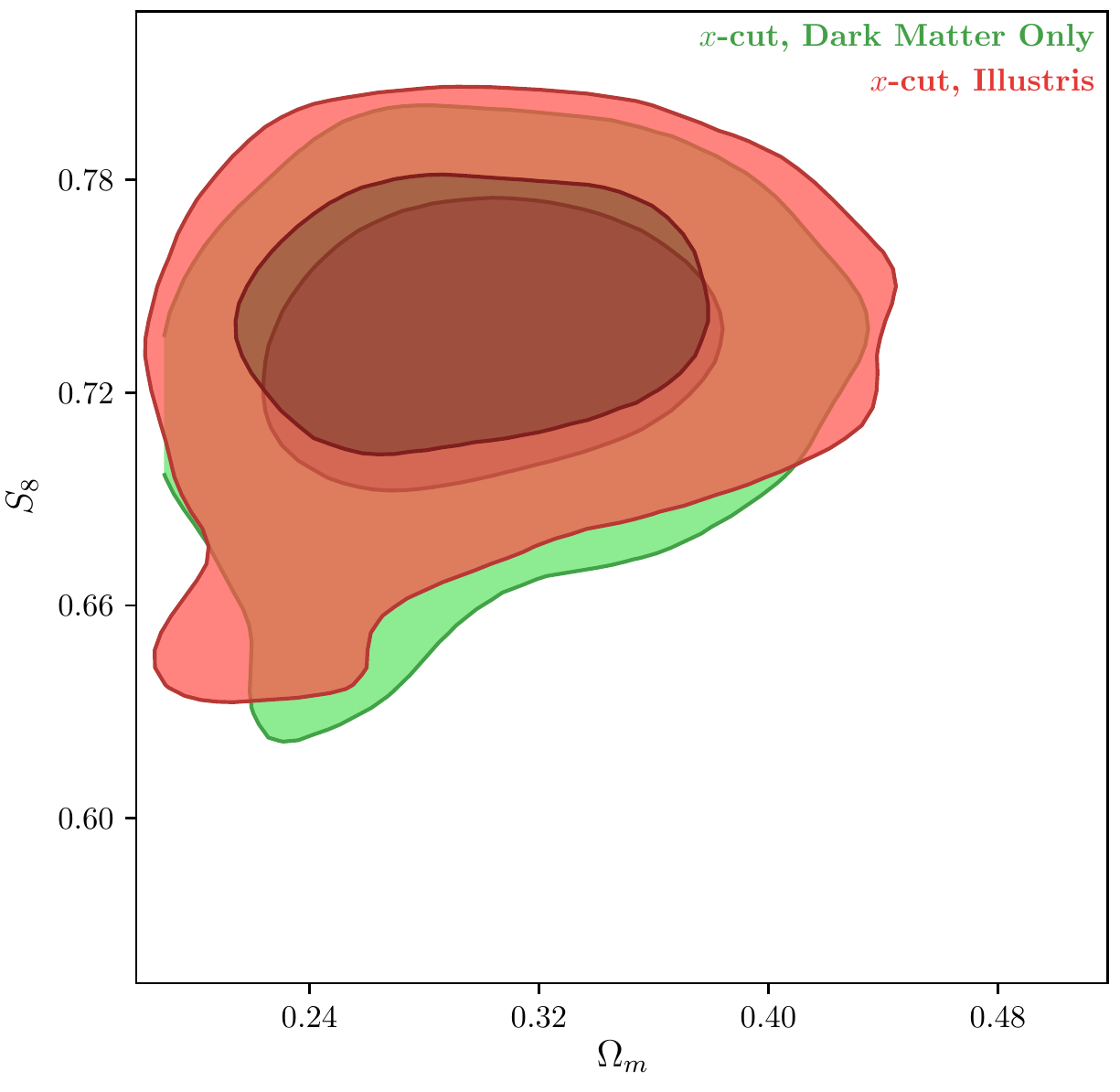}
    \includegraphics[scale = 0.45]{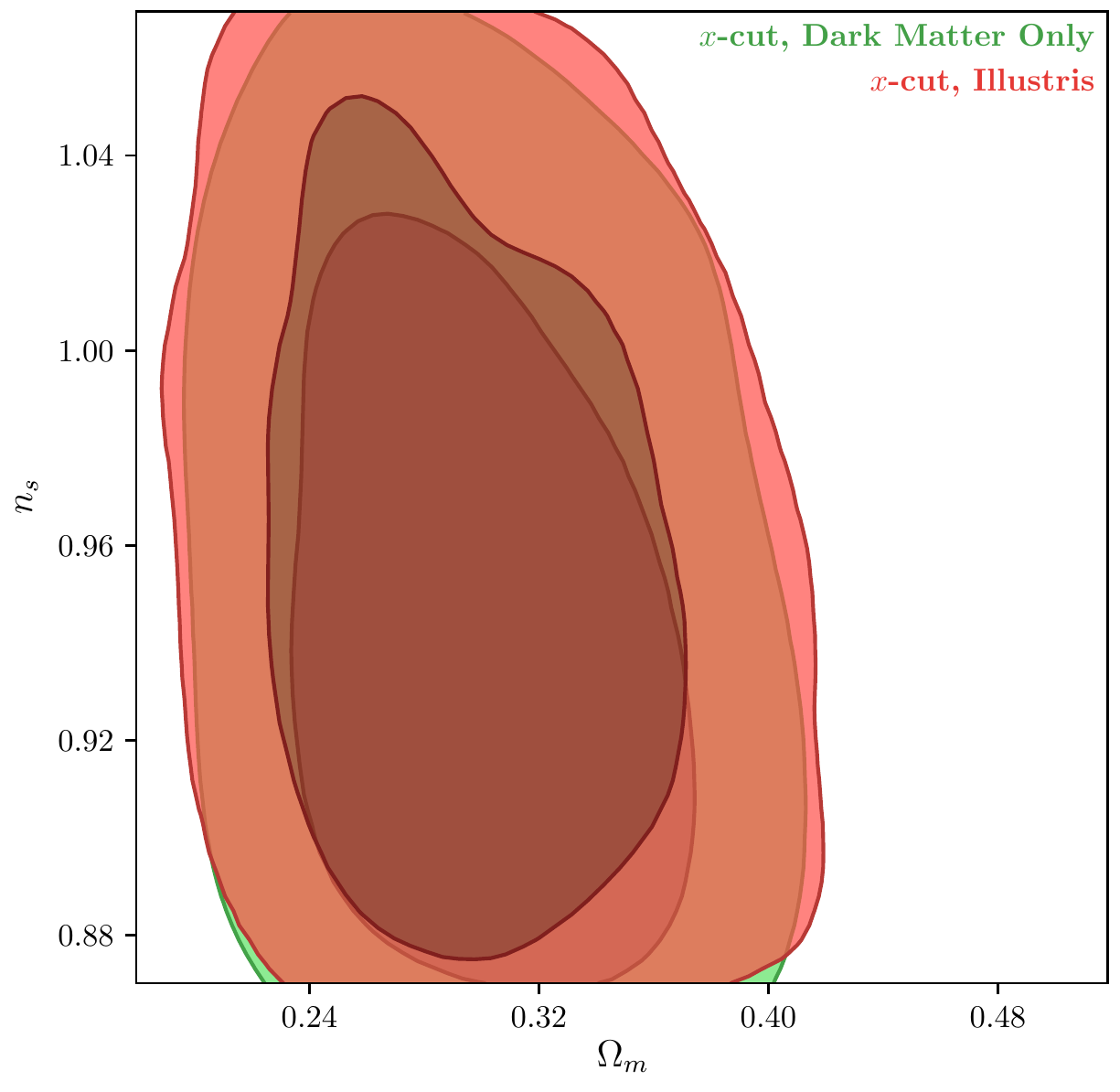}
\includegraphics[scale = 0.45]{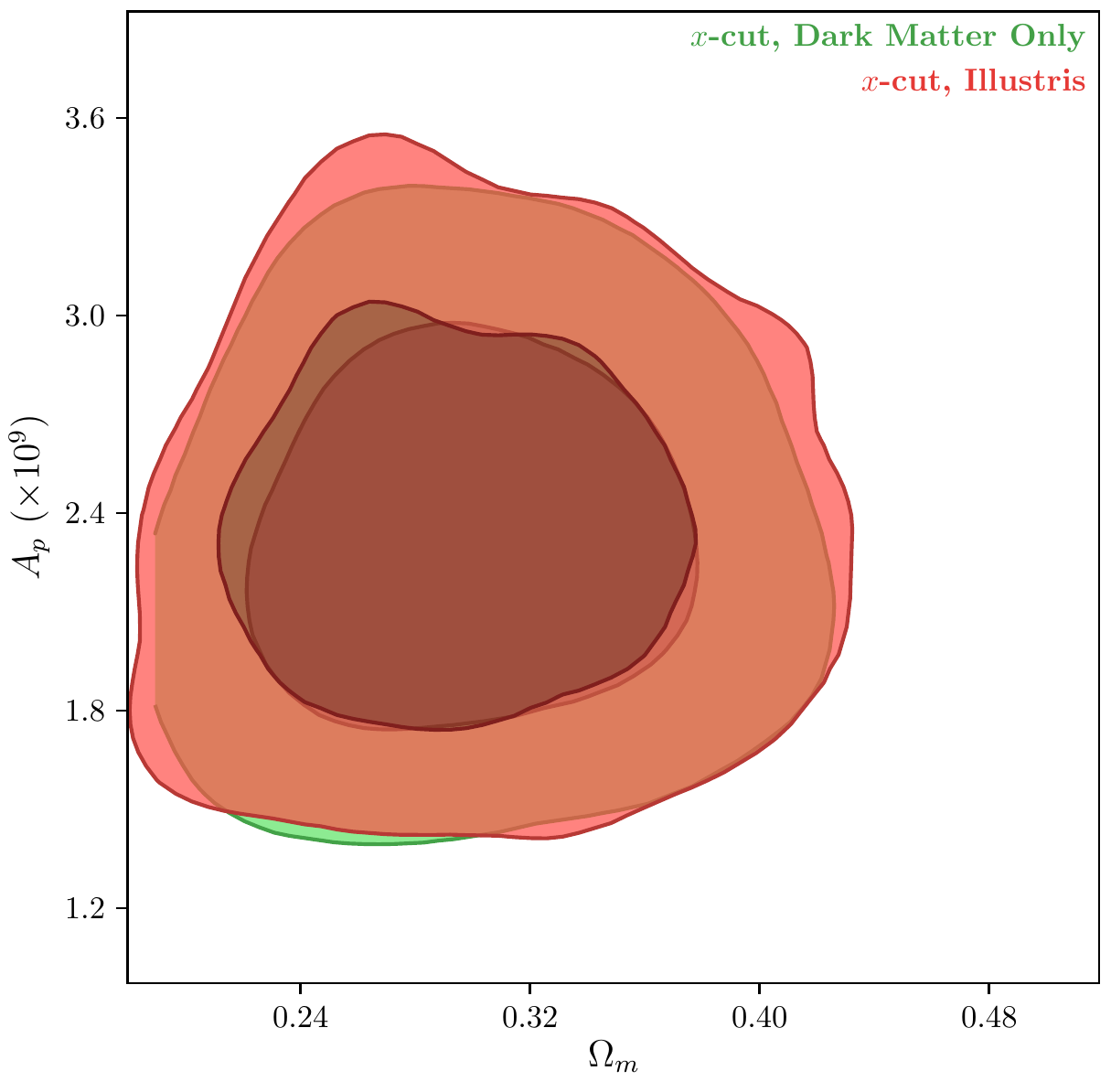}
\caption{{\bf Top:} Parameter constraints from a standard correlation function analysis using the full DESY1 cosmic shear data vector {\bf with no scale cuts} i.e. $20$ angular bins between $2.5$ and $250$ arcmins. We generate the theory vectors using a dark matter only power spectrum and with the Illustris baryonic feedback model. While there is no bias in the amplitude parameter $A_p$, we find discrepancies in $n_s$ and $S_8$. {\bf Bottom:} An $x$-cut cosmic shear analysis. Data points are cut from the analysis when the modelling uncertainty exceeds $5\%$ of the error in the data. This reduces bias in $n_s$ and $S_8$. These constraints are robust to baryonic feedback uncertainties since of all $14$ baryonic feedback models considered in~\cite{huang2018modeling}, Illustris led to the most suppression in power (usually by more than $10\%$ in the range $10 ^{-1} \ h {\rm Mpc} ^{-1} \leq k \leq 10  \ h {\rm Mpc} ^{-1}$). }
\label{fig:comapre_xi_bary}
\end{figure*}

\begin{figure*}[hbt!]
\includegraphics[scale = 0.45]{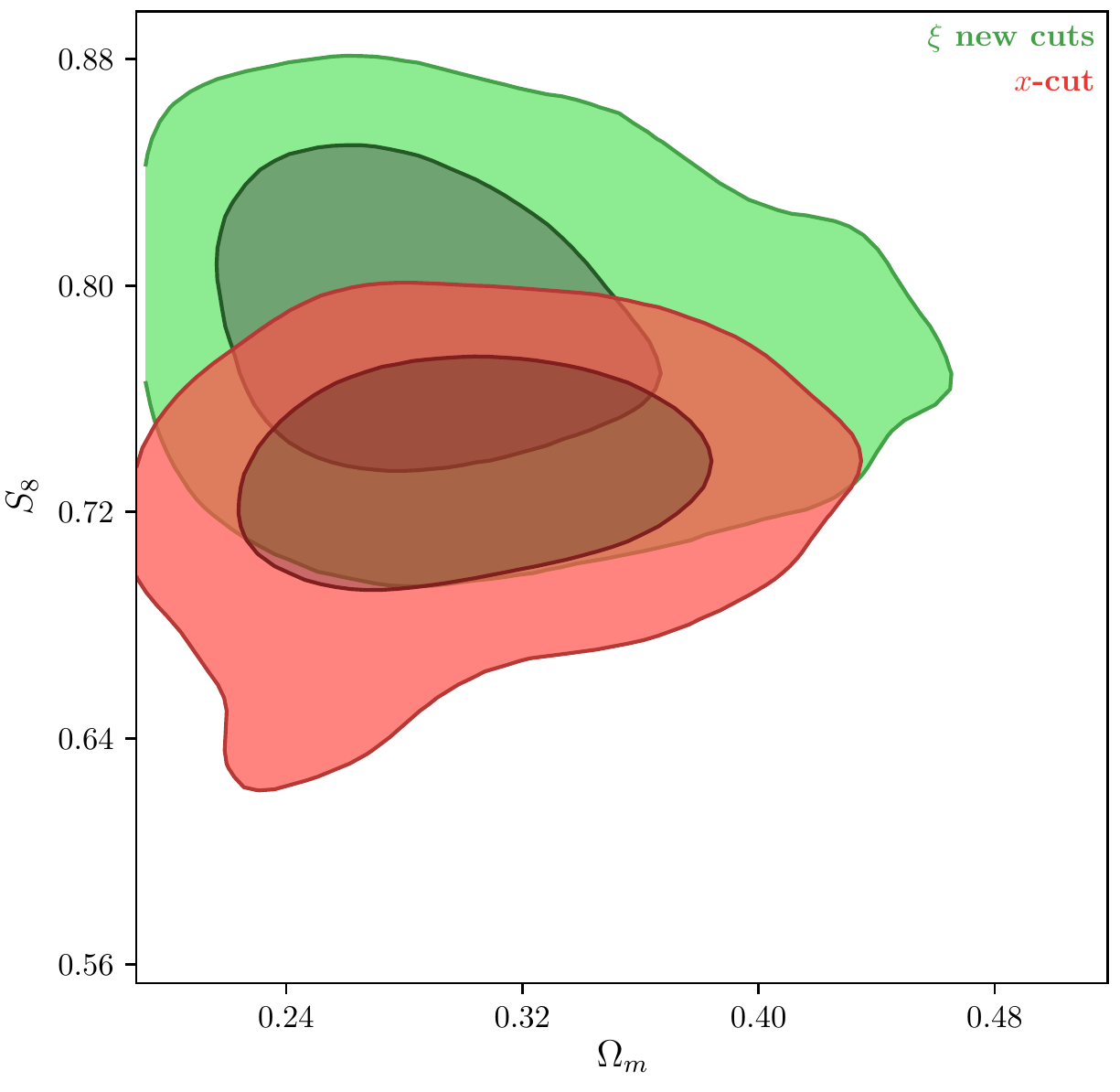}
\includegraphics[scale = 0.45]{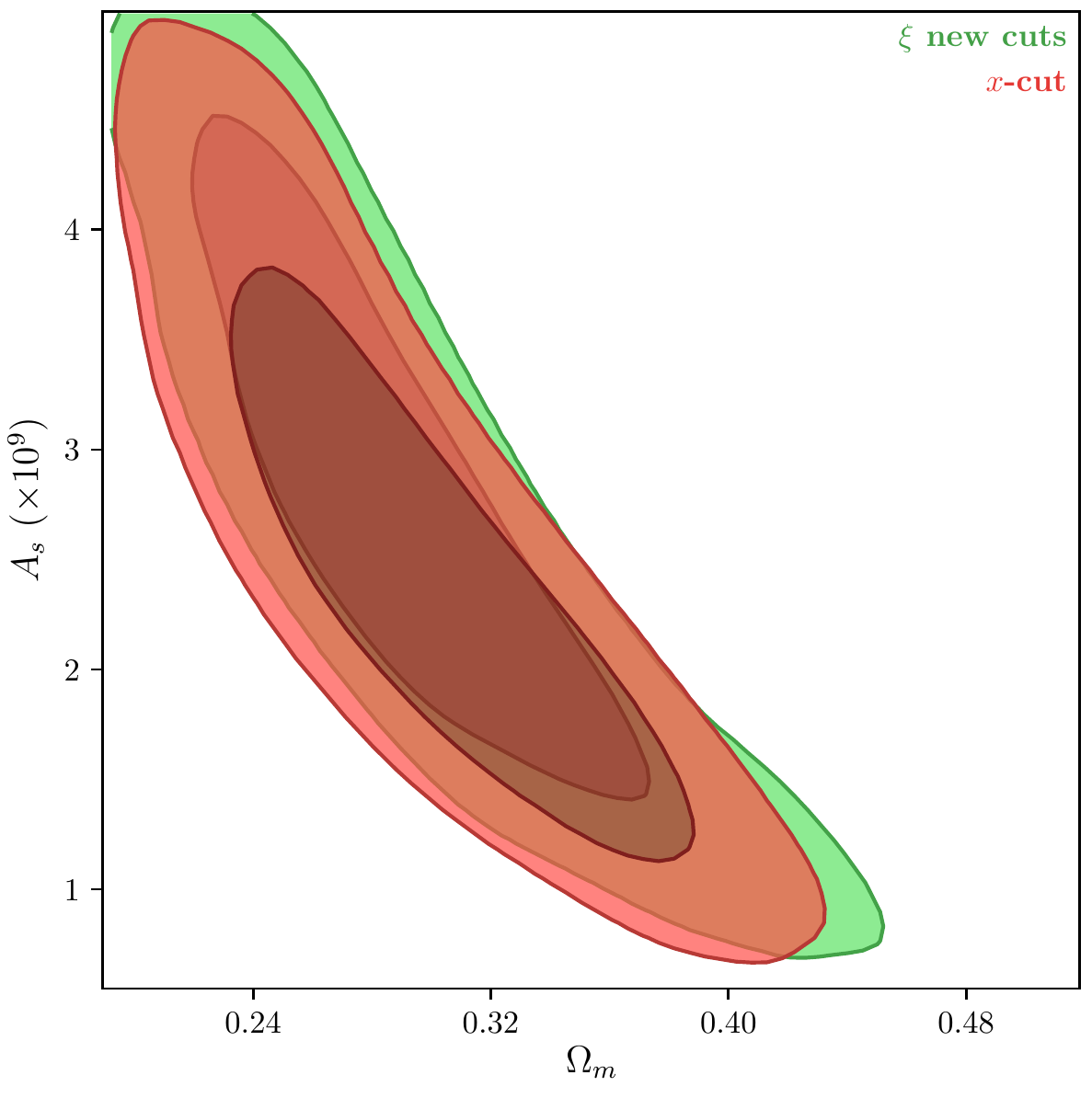}
\includegraphics[scale = 0.45]{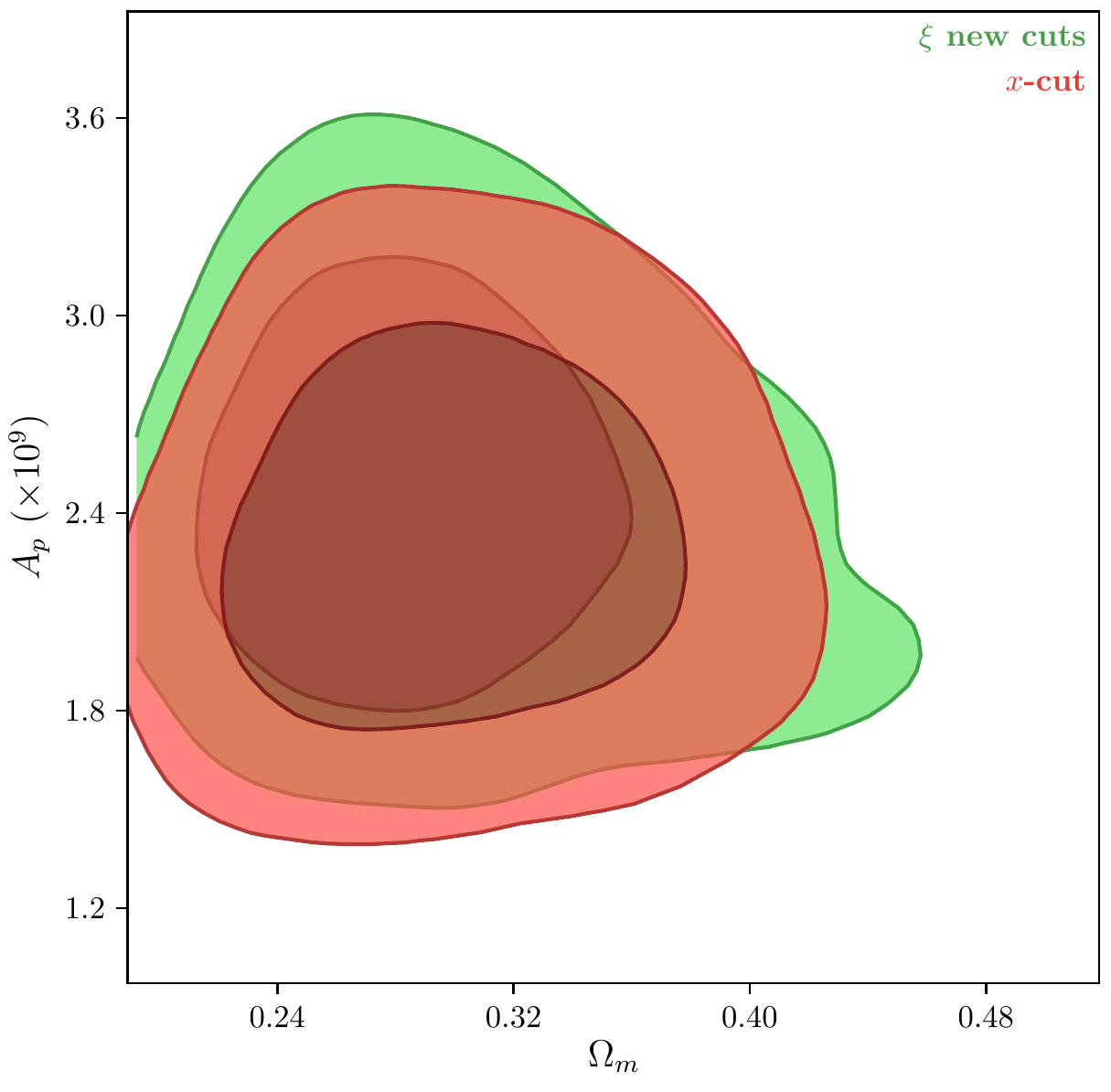}
\caption{ A comparison of the $\xi$ and $x$-cut $68 \%$ and $95 \%$ confidence limits. In both cases scales are cut when the standard deviation of the data point over the four baryonic feedback models considered in this work exceed $5 \%$ of the error, estimated from the covariance. The main difference is the discrepancy in $S_8$. This is driven by the inclusion of small angular scales which must be excluded from the correlation function analysis because of baryonic modelling uncertainty. The symmetric error on $S_8$ is reduced by $32 \%$ when using $x$-cut cosmic shear. Since $x$-cut cosmic shear performs best for deep surveys, with a large number of tomographic bins and precise photo-z's, we expect the relative difference in constraining power between the two methods to become larger in future surveys.}
\label{fig:dm_only}
\end{figure*}

\begin{figure*}[hbt!]
\includegraphics[scale = 0.45]{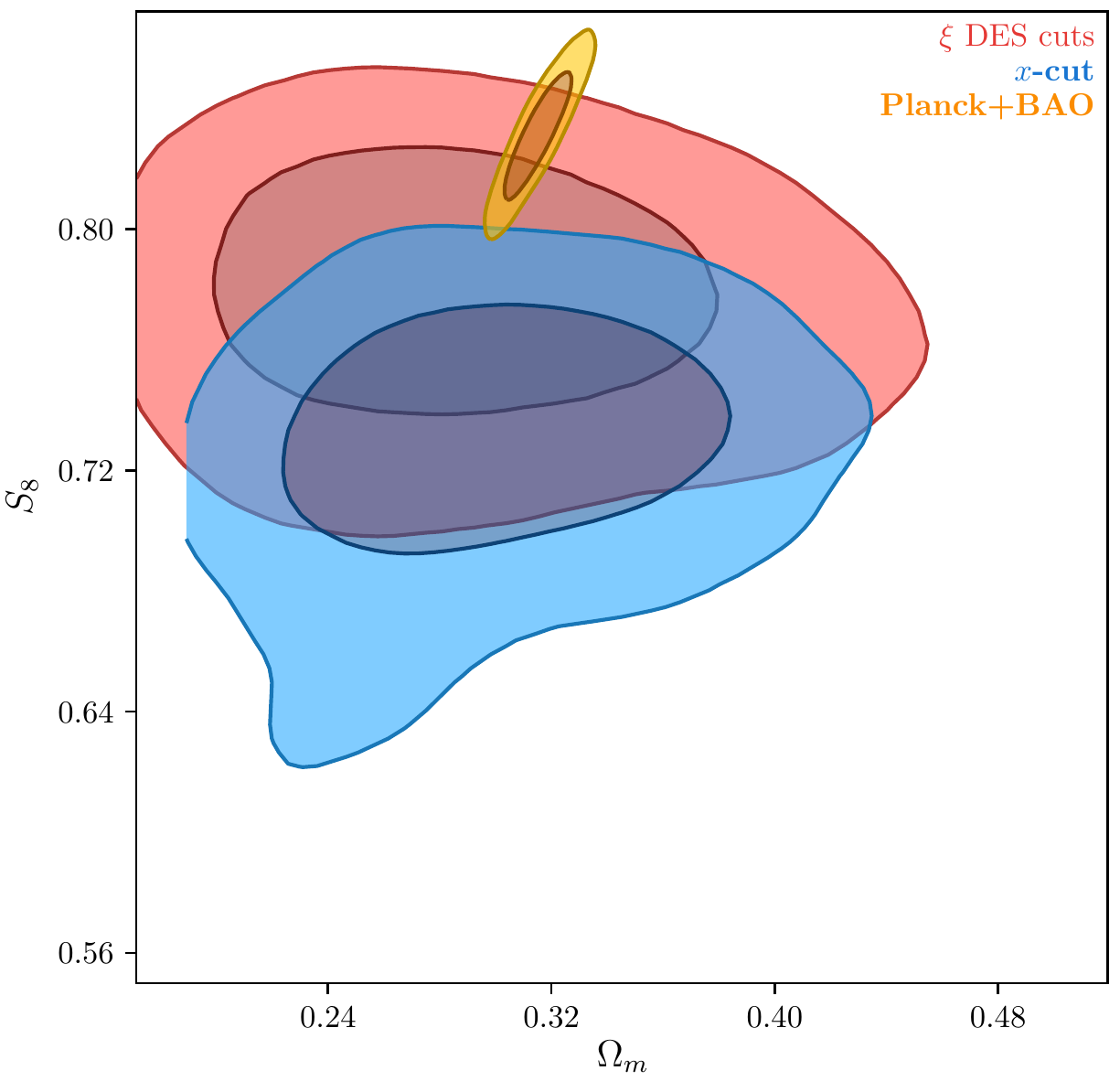}
\includegraphics[scale = 0.45]{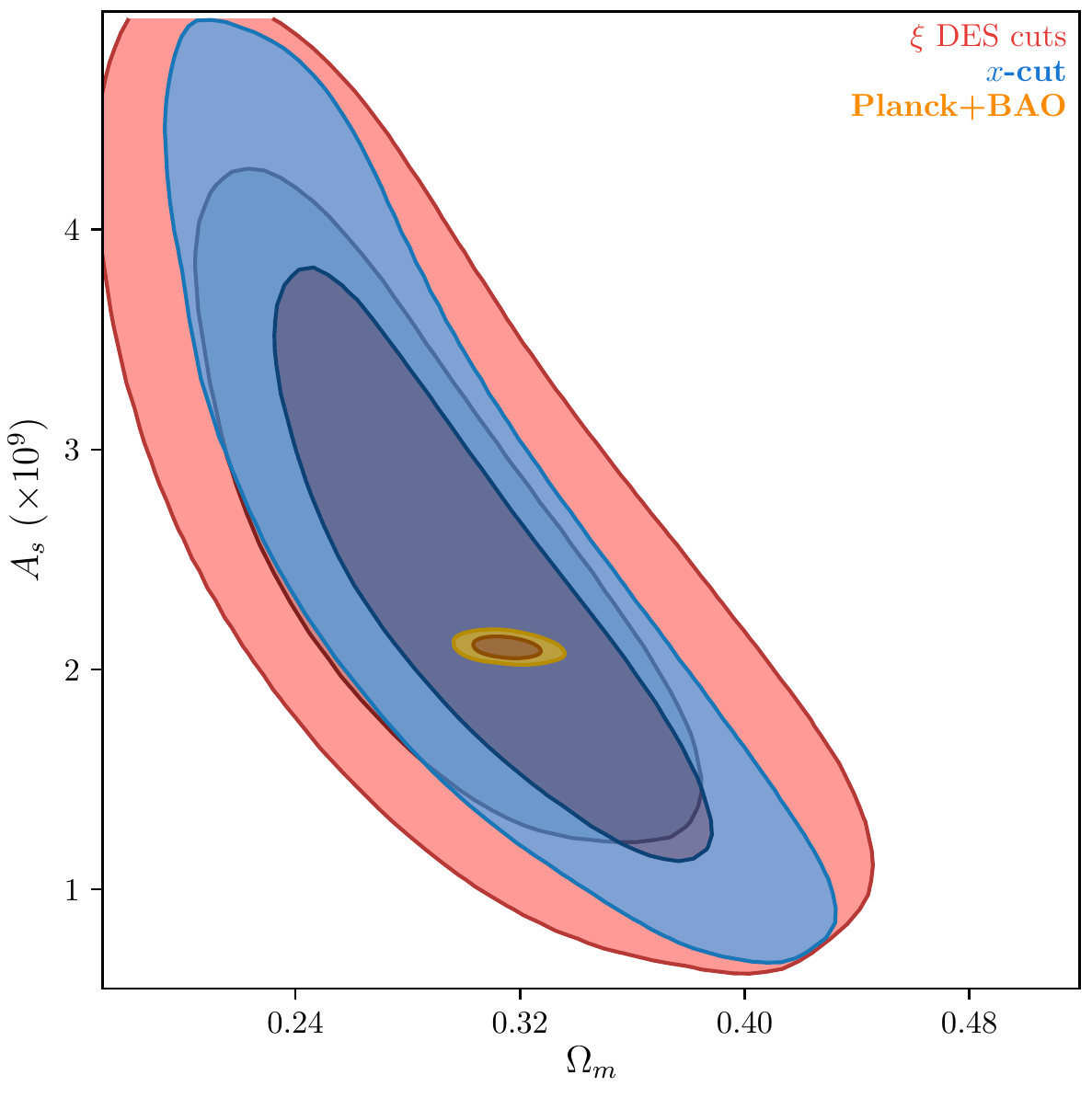}
\includegraphics[scale = 0.45]{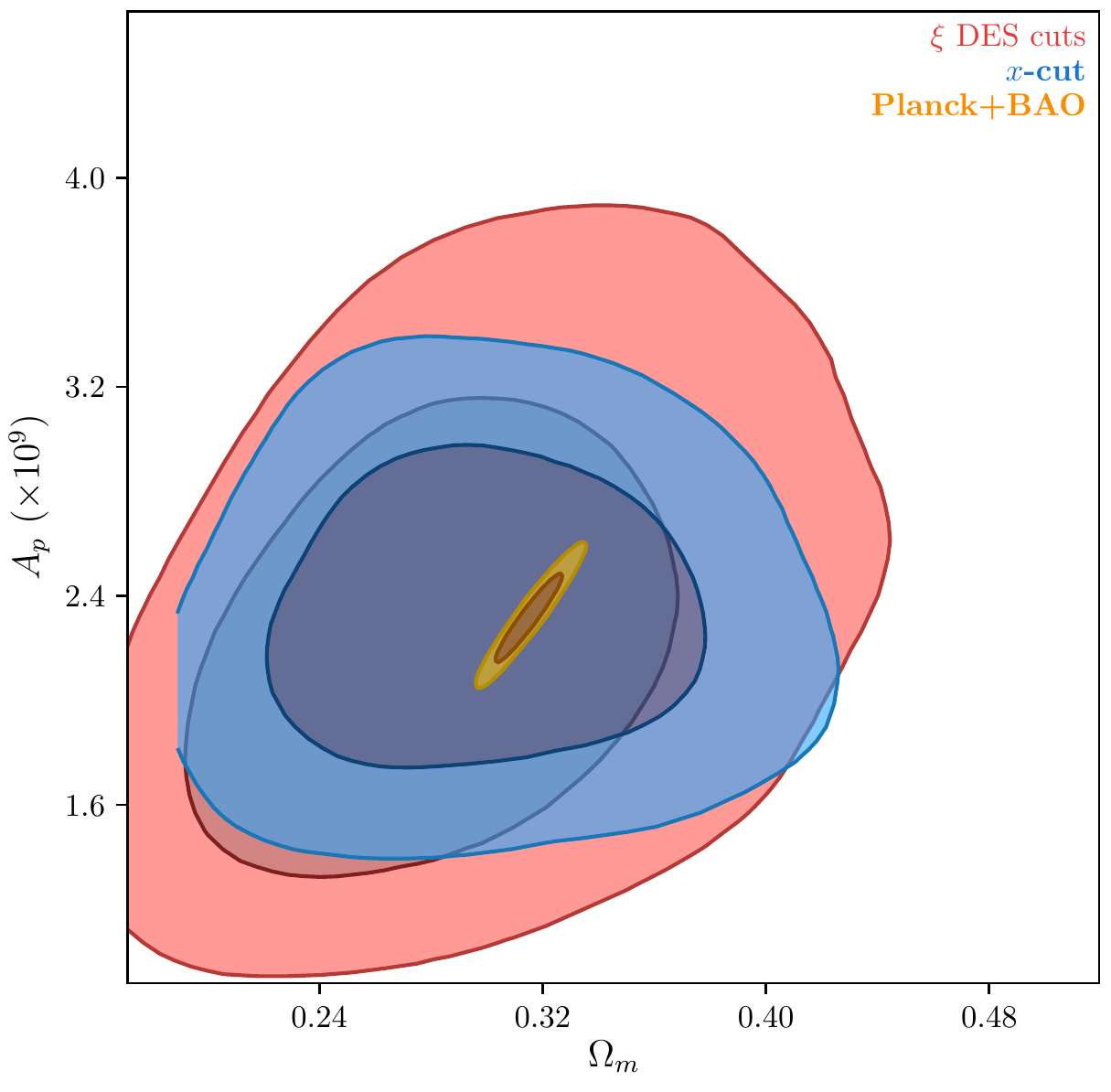}
\caption{A comparison of the $x$-cut cosmic shear constraints against the D18 correlation function fiducial constraints, and the Planck Legacy constraints. {\bf Since the $x$-cut analysis takes a different prior on $h_0$ and uses a different angular scale cut criterion, the reader should refer to Figure~\ref{fig:dm_only} to make a fair comparison between correlation functions and $x$-cut cosmic shear}. Driven by the inclusion of small angular scales in the $x$-cut cosmic shear analysis, the tension with Planck increases to $2.6 \sigma$. It is interesting to note that there is no tension in the primordial amplitudes $A_s$ and $A_p$, so that the $S_8$ tension must be induced through degeneracy with other parameters affecting structure growth in the late Universe.}
\label{fig:dm_only_planck}
\end{figure*}

In this section we compare cosmological constraints from $x$-cut cosmic shear and correlation function analyses. The likelihood is sampled using {\tt Multinest}~\cite{feroz2013importance}. In all cases we assume the same priors as in D18, except we place a Gaussian prior on the Hubble parameter~\cite{heymans2013cfhtlens}, $h_0 \sim \mathcal{N} (0.7, 0.15)$, centred approximately half way between local and high redshift measurements~\cite{riess20162, planck18}. This prior choice ensures the background geometry does not fluctuate over the chain so that $x$-cut shear method removes sensitivity to the desired scales (see the discussion in Section~\ref{sec:formalism}). Cosmic shear in not particularly sensitive to $h_0$, and we check to ensure this does not impact the parameter constraints presented in this work. The priors are summarized in Table~\ref{table:params} for convenience.

\subsection{Verification of the Likelihood Sampled Covariance} \label{sec:verification}

To validate the covariance matrices generated with the likelihood sampling method, we first perform likelihood analyses using the public DESY1 $\xi$-covariance, our recomputed $\xi$-covariance and our derived $x$-cut cosmic shear covariance, with no scale cuts on a simulated data vector. To produce a synthetic correlation function data vector, we draw a random sample from $\mathcal{N} (0, \widehat C_\xi)$, as in the covariance calculation, before adding this to the theoretical expectation at our baseline cosmology. The $x$-cut data vector is produced by taking the BNT transform of the synthetic correlation function data vector. The results are shown in Figure~\ref{fig:cov compare}. We find excellent agreement between the three likelihood chains and we recover the input cosmology confirming the fidelity of the computed covariances.

\subsection{Robustness to Baryonic Modelling Uncertainty} \label{sec:robustness}

Inaccurate models of baryonic physics can lead to biased parameter estimates -- even at the precision of today's experiments. To see this, we perform two correlation function likelihood analyses of the DESY1 data, without taking any scale cuts. In the first chain we use a dark matter only power spectrum and in the second, we use the Illustris baryonic feedback model. The resulting parameter constraints are shown in the top row of Figure~\ref{fig:comapre_xi_bary}. Although the two models give very similar constraints in the amplitude parameter $A_p$, we find biases in $n_s$ and $S_8$. Despite being poorly constrained, $n_s$ is particularly affected because baryonic feedback suppresses high $k$-modes more than low $k$-modes, changing the inferred tilt of the matter power spectrum.
\par This bias is typically avoided by taking conservative angular scale cuts, but as we have argued this also removes useful information. In D18, data points were excluded from the analysis if the difference between a dark matter only model and the OWLS-AGN model exceeded $2\%$. 
\par Next we perform an $x$-cut cosmic shear analysis. This time we conservatively remove data points where the
standard deviation between the four baryonic feedback models, $\sigma_{\rm sim}$, is less than $5\%$ of the error on the data, $\sigma_{\rm data}$, where $\sigma_{\rm data}$ is computed by taking the square root of the diagonal of the covariance matrix so that: $\sigma_{\rm sim} / \sigma_{\rm data} < 0.05$, corresponding to the grey shaded region in Figure~\ref{fig:plus}.
\par Parameter constraints inferred using a dark matter only model and the Illustris feedback model are compared in Figure~\ref{fig:comapre_xi_bary}. Despite removing an extremely limited number of data points, the $x$-cut cosmic shear technique removes the majority of the bias between the Illustrius and dark matter only $n_s$ and $S_8$  constraints. The remaining bias is due to the fact that cut scales were chosen by considering 4 baryonic feedback models, not just Illustrius. Given that of the 14 baryonic feedback models consider in~\cite{huang2018modeling}, the Illustris model is by far the most extreme, with the largest suppression in power, the bias using other baryonic feedback models will be much smaller.
\par We take the dark matter only $x$-cut analysis as our fiducial case. In the next section we compare these results to the D18 constraints and a correlation function analysis with the same scale cut criterion as before.

\subsection{Parameter Constraints}

\begin{table*}
    \centering
    \caption{Final parameter constraints. $S_8$ is computed using the dark matter only power spectrum from {\tt Halofit}. All error bars in this paper are computed using {\tt ChainConsumer} with no kernel density estimation (KDE). Instances where {\tt ChainConsumer} fails to compute errors are indicated with an NC. Legacy results are displayed below the dividing line at the middle of the table.}
    \label{tab:example}
    \begin{tabular}{ccccc}
        \hline
		Model & $S_8$ & $\Omega_m$ & $A_s$ $(\times 10 ^9)$ & $A_p$ $(\times 10 ^9)$ \\ 
		\hline
		$x$-cut & $0.737^{+0.022}_{-0.029}$ & $0.290^{+0.054}_{-0.044}$ & $\left( 21.5^{+9.1}_{-7.3} \right) \times 10^{-10}$ & $\left( 22.4^{+4.6}_{-3.3} \right) \times 10^{-10}$ \\ 
		$\xi$ new cuts & $0.802^{+0.020}_{-0.048}$ & $0.288^{+0.028}_{-0.059}$ & $\left( 26.9^{+9.3}_{-11.2} \right) \times 10^{-10}$ &  NC \\ 
		$\xi$ DES cuts & $0.781^{+0.028}_{-0.027}$ & $0.255^{+0.070}_{-0.035}$ & NC & $\left( 20.7^{+6.4}_{-4.7} \right) \times 10^{-10}$ \\ 
		Planck+BAO & $0.831^{+0.013}_{-0.012}$ & $\left( 315.9^{+6.3}_{-8.4} \right) \times 10^{-3}$ & $\left( 210.1^{+2.8}_{-3.3} \right) \times 10^{-11}$ & $\left( 23.1\pm 1.0 \right) \times 10^{-10}$ \\ 
		\hline
    \end{tabular}
\end{table*}

\par In Figure~\ref{fig:dm_only} we compare our fiducial $x$-cut cosmic shear and correlation function parameter constraints. In both cases we have used the criterion from the last section, and cut all scales where the standard deviation of the data point over the four baryonic feedback models exceeds $5\%$ of the error (estimated from the covariance). The resulting confidence regions are very similar, except in $S_8$. Here the $x$-cut cosmic shear constraints are shifted down by $\sim 1 \sigma$ and the symmetric error is $32 \%$ smaller. This is driven by the inclusion of small angular scales which must be cut from the correlation function analysis. It is worth remembering that while this is a very modest improvement in constraining power, this is expected to improve in future surveys as $x$-cut cosmic shear works better relative to correlation functions for deep surveys, with a large number of tomographic bins and precise photo-z's.
\par In Figure~\ref{fig:dm_only_planck} we plot our fiducial $x$-cut constraints against the D18 results and the Planck Legacy (TT,TE,EE+lowE + Baryonic Acoustic Osciallations (BAO)) confidence regions~\cite{planck18}. $x$-cut cosmic shear places noticeably tighter constraints on all parameters and the symmetric error on $S_8$ is reduced by $9 \%$. As we have shown, these results are robust to baryonic modelling uncertainty. Driven by the inclusion of small angular scales in the $x$-cut cosmic shear analysis, the tension with Planck increases to $2.6 \sigma$. This is similar to the COSEBI constraints presented in~\cite{asgari2020kids+} which also includes information from small angular scales while remaining robust to baryons. It is interesting to note that there is no tension in the primordial amplitudes $A_s$ and $A_p$, so that the $S_8$ tension must be induced through degeneracies with other parameters affecting structure growth in the late Universe.

\begin{figure*}[hbt!]
\includegraphics[scale=1.35]{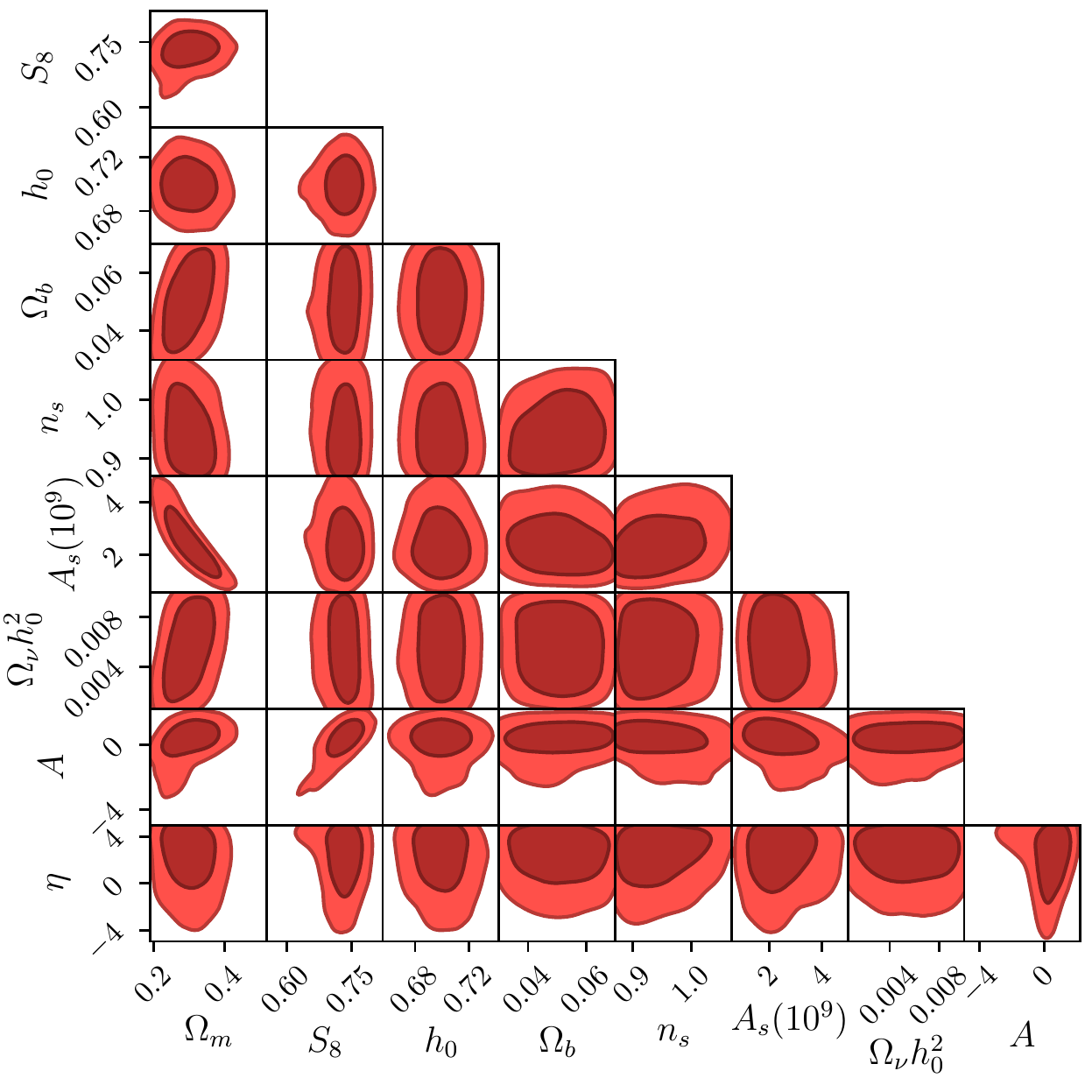}
\caption{Cosmological parameter constraints from the fiducial $x$-cut analysis, after marginalizing over the redshift biases, $\Delta z ^i$, and the multiplicative biases, $m^i$.}
\label{fig:full}
\end{figure*}

\par A summary of the parameter constraints discussed in this section are given in Table~\ref{tab:example} and the fiducial $x$-cut analysis cosmological parameter constraints are shown in Figure~\ref{fig:full}.

\section{Conclusions and Future Prospects} \label{sec:conclusions}
We have presented a new method called $x$-cut cosmic shear. This technique optimally removes sensitivity to small poorly modeled scales and is the configuration space analog of $k$-cut cosmic shear~\cite{taylor2018k-cut}.
\par We show how to compute the $x$-cut covariance matrix from a correlation function covariance matrix in a few minutes on a single CPU. This method could be used to compute a $k$-cut covariance matrix from a $C_\ell$ covariance matrix.
\par Using our derived covariance matrix, we perform an $x$-cut cosmic shear likelihood analysis of the DESY1 shear data. Since the information has been sorted by scale, we take more aggressive cuts than would be possible in a correlation function analysis, tightening constraints on $S_8$ relative to the correlation function analysis. By comparing parameter constraints found using an extreme baryonic physics model to the dark matter only case, we ensure our results are robust to baryonic modelling uncertainties. 
\par The tension with Planck and BAO measurements in $S_8$ increases to $2.6 \sigma$. This is driven by the inclusion of data at small angular scales but could be due to previously unexplored systematics. It is worth noting that the photometric redshifts in this analysis are calibrated on $30$ photometric band COSMOS data and previous studies find that the tension with Planck increases if the photo-zs are calibrated on spectroscopic data~\cite{hildebrandt2020kids+, joudaki2019kids+}.
\par As photometric redshift estimation and the number of tomographic bins increases, the performance of the method will actually improve. This is because the BNT reweighted lensing efficiency kernels, $\tilde q ^ i (\chi)$, will have less overlap. In future surveys, efforts should be made to estimate, $\xi_\pm (\theta)$ down to very small angular scales. $x$-cut cosmic shear will enable the extraction of useful information from these scales. 
\par Even after the BNT transformation, constraining power is degraded by the scale cut. This transformation just makes the cut less suboptimal than a traditional $C_\ell$ or $\xi_\pm$ analysis. For this reason it remains important to model the baryonic physics as accurately as possible. 
\par Calibrating the baryonic feedback models using external observations~\cite{debackere2020impact, van2020exploring} is a promising approach. But ultimately, some combination of baryonic feedback mitigation strategies (see e.g ~\cite{eifler2015accounting, huang2018modeling, mead2015accurate, harnois2015baryons}), improved modelling and $x$/$k$-cut cosmic shear will likely be warranted. 
\par Beyond helping alleviate baryonic modelling uncertainties, $x$-cut cosmic shear will help constrain theories of modified gravity. Although it may be computationally infeasible to emulate the matter power spectrum at percent-level accuracy down $k \sim 10 \ h {\rm Mpc}^{-1}$, for {\it all} theories, our method ensures that information lost to cut scales will at least be minimal. 
\par Compared to correlation functions, the $x$-cut cosmic shear method (and $x$-cut cosmic shear galaxy-galaxy lensing in $3 \times 2$-point analyses) has multiple advantages. We have shown this comes at virtually no additional additional computational cost -- easily slotting into existing pipelines. For these reasons we advocate for the use of $x$-cut cosmic shear in upcoming surveys.

\section{Acknowledgements}
 PLT thanks Agn\`es Fert\'e, Jason Rhodes and Tom Kitching for useful conversations. We are indebted to Hung-Jin Huang for making the baryonic feedback data publicly available. PLT acknowledges support for this work from a NASA Postdoctoral Program Fellowship. Part of the research was carried out at the Jet Propulsion Laboratory, California Institute of Technology, under a contract with the National Aeronautics and Space Administration. The authors would like to thank the anonymous referee whose comments have significantly improved the manuscript.

\par This project used public archival data from the Dark Energy Survey (DES). Funding for the DES Projects has been provided by the U.S. Department of Energy, the U.S. National Science Foundation, the Ministry of Science and Education of Spain, the Science and Technology Facilities Council of the United Kingdom, the Higher Education Funding Council for England, the National Center for Supercomputing Applications at the University of Illinois at Urbana-Champaign, the Kavli Institute of Cosmological Physics at the University of Chicago, the Center for Cosmology and Astro-Particle Physics at the Ohio State University, the Mitchell Institute for Fundamental Physics and Astronomy at Texas A\&M University, Financiadora de Estudos e Projetos, Funda{\c c}{\~a}o Carlos Chagas Filho de Amparo {\`a} Pesquisa do Estado do Rio de Janeiro, Conselho Nacional de Desenvolvimento Cient{\'i}fico e Tecnol{\'o}gico and the Minist{\'e}rio da Ci{\^e}ncia, Tecnologia e Inova{\c c}{\~a}o, the Deutsche Forschungsgemeinschaft, and the Collaborating Institutions in the Dark Energy Survey.

The Collaborating Institutions are Argonne National Laboratory, the University of California at Santa Cruz, the University of Cambridge, Centro de Investigaciones Energ{\'e}ticas, Medioambientales y Tecnol{\'o}gicas-Madrid, the University of Chicago, University College London, the DES-Brazil Consortium, the University of Edinburgh, the Eidgen{\"o}ssische Technische Hochschule (ETH) Z{\"u}rich,  Fermi National Accelerator Laboratory, the University of Illinois at Urbana-Champaign, the Institut de Ci{\`e}ncies de l'Espai (IEEC/CSIC), the Institut de F{\'i}sica d'Altes Energies, Lawrence Berkeley National Laboratory, the Ludwig-Maximilians Universit{\"a}t M{\"u}nchen and the associated Excellence Cluster Universe, the University of Michigan, the National Optical Astronomy Observatory, the University of Nottingham, The Ohio State University, the OzDES Membership Consortium, the University of Pennsylvania, the University of Portsmouth, SLAC National Accelerator Laboratory, Stanford University, the University of Sussex, and Texas A\&M University.
\bibliographystyle{apsrev4-1.bst}
\bibliography{bibtex.bib}

\end{document}